\documentclass[review,onefignum,onetabnum]{siamonline250211}

\nolinenumbers
\usepackage{lipsum}
\usepackage{amsfonts}
\usepackage{graphicx}
\usepackage{epstopdf}
\usepackage{algorithmic}
\usepackage{amsopn}
\usepackage{wrapfig}
\usepackage{sidecap}

\ifpdf
  \DeclareGraphicsExtensions{.eps,.pdf,.png,.jpg}
\else
  \DeclareGraphicsExtensions{.eps}
\fi

\newsiamremark{remark}{Remark}
\newsiamremark{hypothesis}{Hypothesis}
\crefname{hypothesis}{Hypothesis}{Hypotheses}
\newsiamthm{claim}{Claim}

\headers{Hodge Decomposition Guides the Optimization of Synchronization}{C. Purple, P.S. Skardal, and D. Taylor}

\title{Hodge Decomposition Guides the Optimization of Synchronization over Simplicial Complexes\thanks{
\funding{This work was funded in part by National Science Foundation grant DMS-2401276. Any opinions, findings, and conclusions or recommendations expressed in this material are those of the author(s) and do not necessarily reflect the views of the NSF.}}}

\author{Cameron Purple\thanks{Department of Mathematics, SUNY Buffalo, Buffalo, NY 
  (\email{cz22@buffalo.edu}).} \and
  Per Sebastian Skardal\thanks{Department of Mathematics, Trinity College, Hartford, CT 
  (\email{persebastian.skardal@trincoll.edu}).}
\and Dane Taylor\thanks{School of Computing and Department of Mathematics \& Statistics, University of Wyoming, Laramie, WY
  (\email{dane.taylor@uwyo.edu}).}}

\makeatletter
\newcommand*{\addFileDependency}[1]{
  \typeout{(#1)}
  \@addtofilelist{#1}
  \IfFileExists{#1}{}{\typeout{No file #1.}}
}
\makeatother

\begin{document}
\maketitle

\begin{abstract}
    Despite  growing interest in synchronization dynamics over ``higher-order'' network models, optimization theory for such systems is limited.
    Here, we study a family of Kuramoto models  inspired by algebraic topology in which oscillators are coupled over simplicial complexes (SCs) using their associated Hodge Laplacian matrices.
    We optimize such systems by extending the synchrony alignment function---an optimization framework for synchronizing graph-coupled heterogeneous oscillators.
    Computational experiments are given to illustrate how this approach can effectively solve a variety of combinatorial problems including the joint optimization of projected synchronization dynamics onto lower- and upper-dimensional simplices within SCs.
    %
    We also investigate the role of SC homology and develop bifurcation theory to characterize the extent to which optimal solutions are contained within (or spread across) the three Hodge subspaces. 
    Our work extends optimization theory to the setting of higher-order networks, provides practical algorithms for Hodge-Laplacian-related dynamics including (but not limited to) Kuramoto oscillators, and paves the way for an emerging field that interfaces algebraic topology, combinatorial optimization, and dynamical systems.
\end{abstract}



\section{Introduction}

Synchronization is a self-organized phenomenon in complex systems in which a set of dynamical systems are coupled together and collectively obtain similar states (e.g., frequencies, phases, vector headings, etc.) \cite{pikovsky2001universal}. It is vital to our understanding for diverse applications ranging from biological processes (e.g., neurons \cite{fries2005mechanism, breakspear2010generative},  cardiac pacemaker cells \cite{glass1988clocks}, and firefly populations \cite{buck1988synchronous}) to   engineered systems (e.g.,  power grids \cite{nishikawa2015comparative, guo2021overviews}, Josephson junctions  \cite{wiesenfeld1998frequency}, and  structural instabilities of bridges \cite{strogatz2005crowd}).  Among the many models for synchronization,  Kuramoto  phase-oscillator models \cite{kuramoto2003chemical} have become widely investigated due to their relevance to weakly coupled dynamical systems \cite{strogatz2000kuramoto,nakao2016phase} and their ability to allow analytical tractability \cite{strogatz2000kuramoto,acebron2005kuramoto,ott2008low}. In particular, there is a well-developed literature studying Kuramoto oscillators coupled over sparse graphs---i.e., oscillators are assigned to vertices/nodes and pairwise interactions between oscillators are assigned to edges \cite{restrepo2005onset, restrepo2006synchronization}.  

For diverse applications, it is important to consider  systems that are optimized to enhance synchronization, which can arise  by strategic engineering (or in the case of biology, natural selection and evolution). In principle, one can optimize a network of Kuramoto phase oscillators  by tuning dynamical properties (e.g., oscillators' natural frequencies) or by restructuring the network of interactions \cite{nishikawa2006synchronization, donetti2005entangled,brede2008synchrony}. In this work, we focus on extending the synchrony alignment function (SAF) \cite{skardal2014optimal}, which is a popular optimization framework for network-coupled systems  of heterogeneous phase oscillators. 
The SAF was originally developed for undirected graphs and has  been extended to directed graphs \cite{skardal2016optimal}, systems with uncertainty  \cite{skardal2019synchronization}, modular structure \cite{chamlagai2022grass}, and higher-order interactions \cite{skardal2021higher}.

Notably, there is growing interest to extend research on network-coupled dynamical systems beyond graph-based modeling by utilizing higher-order combinatorial objects (e.g., simplicial complexes and hypergraphs) that can implement multiway interactions  \cite{battiston2021physics,bianconi2021higher,majhi2022dynamics}. One advantage of simplicial complexes (SCs) is their connection to the mathematical field of topology, and therefore SC-based models open up new applications for topological tools including homology and Hodge theory. For example, a Hodge Laplacian matrix can be used to encode interactions  among $d$-dimensional simplices (e.g., 1-dimensional `lines' or edges) that connect to common neighboring simplices of lower- and upper-dimensions (e.g., are connected through neighboring 0-dimensional `vertices' and 2-dimensional `triangles').  These matrices have enabled SC-based extensions for dynamics including random walks \cite{schaub2020random}, consensus~\cite{ziegler2022balanced}, signal processing \cite{barbarossa2020topological, ghorbanchian2021higher}, and  neural networks for machine learning \cite{roddenberry2019hodgenet} (which are subsequently emerging new application areas for applied topology).

\begin{figure}[!t]
\centering
    \includegraphics[width=\linewidth]{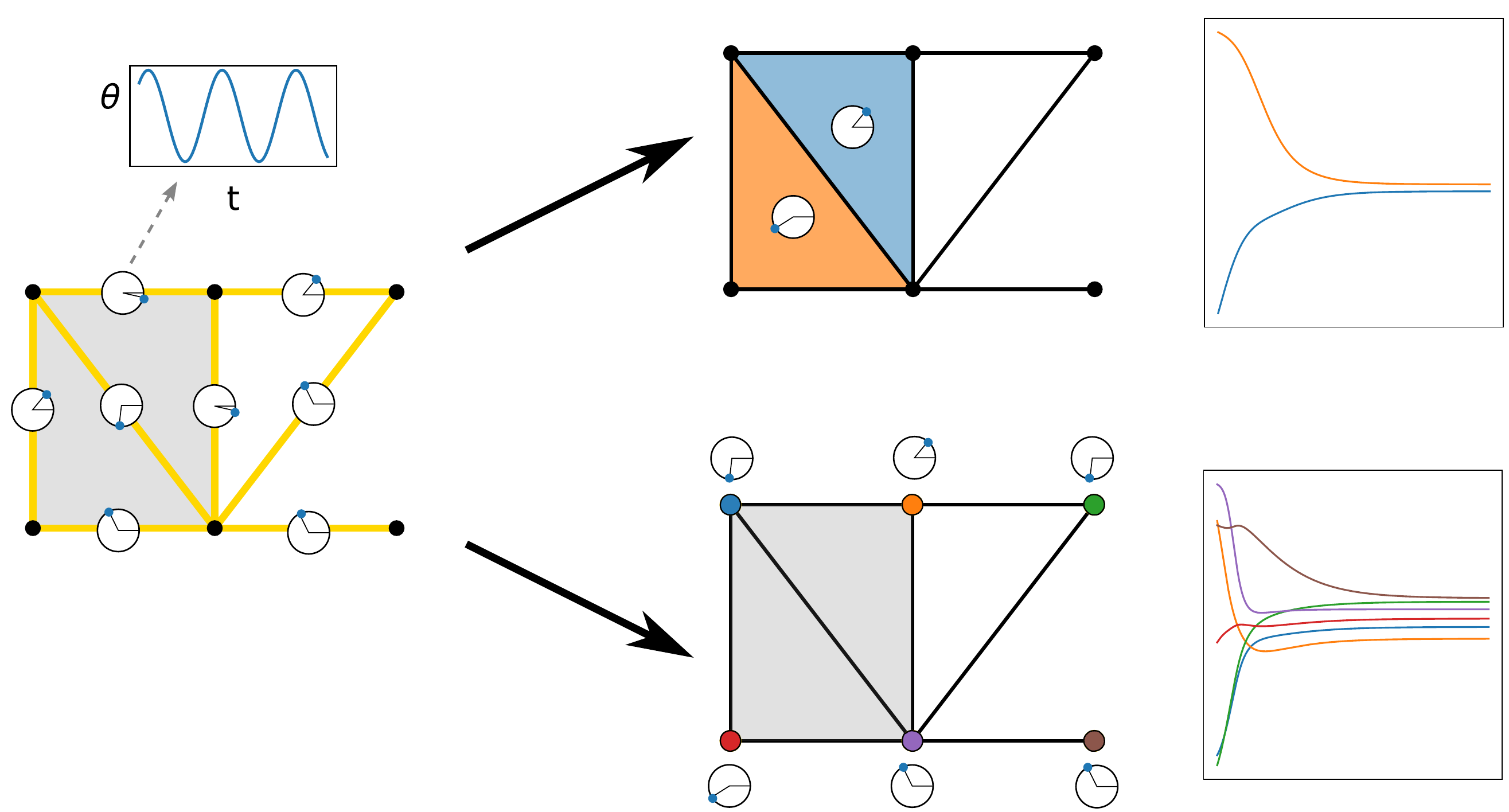}
    \vspace{-.3cm}
    \caption{{\bf Hodge Laplacian Kuramoto (HLK) phase-oscillator dynamics give rise to two simultaneous dynamical systems obtained through lower- and upper-dimensional projections.} We study HLK dynamics (see Sec.~\ref{sec:HodgeLaplacian}) in which phase oscillators are assigned to  1-simplices (i.e., edges) within a simplicial complex (SC). The associated dynamics can be projected onto their neighboring 0-simplices (i.e., nodes) and 2-simplices (i.e., filled-in triangles), both of which can exhibit synchronization and phase locking.
    }
  \label{fig:projections}
\end{figure}

Herein, we extend the literature studying Kuramoto oscillators coupled through a SC \cite{skardal2021higher,ghosh2025transitions,nurisso2024unified,carletti2023global}. We focus on a Hodge Laplacian Kuramoto  (HLK) model \cite{nurisso2024unified,carletti2023global} that extends the Kuramoto model to SCs and is inspired by Hodge Laplacians and algebraic topology.
We introduce and analyze a generalized HLK dynamics that incorporates our prior work on Balanced Hodge Laplacians \cite{ziegler2022balanced}, in which   a ``balancing parameter'' was introduced to tune the relative strength of interactions through lower- and upper-dimensional simplices.
%
%
Interestingly, HLK models can give rise to two simultaneous synchronization processes that occur for projected dynamical systems that result after considering a dynamical system defined over $k$-dimensional simplices and projecting the dynamics onto their neighboring simplices with dimensions $k\pm1$, as shown in Fig.~\ref{fig:projections}.

%
%

Our main contribution is the development of a theoretical optimization framework that can effectively maximize Kuramoto order parameters for these projected dynamical systems. Specifically, we extend the SAF framework to HLK models, thereby complementing prior extensions of the SAF to higher-order networks \cite{skardal2021higher}. 
We highlight our framework's performance by posing and solving (either exactly or approximately) combinatorial optimization problems that seek to enhance synchronization  by optimally choosing  the oscillators' natural  frequencies for a fixed  SC.
We also investigate the role of SC homology and characterize the extent to which the optimal frequency vectors relate to the three Hodge subspaces (e.g., the harmonic, gradient, and curl subspaces). 
Among other results, we find that synchronization is strongest when the natural frequencies lie within the harmonic subspace---which we interpret as providing a `safe haven' for frequency heterogeneity that does not contribute to the projected dynamics. 
To further investigate the role of SC homology in shaping synchrony-optimal systems, we introduce and study spectrally constrained optimization problems and develop bifurcation theory to characterize when optimal frequency vectors are contained within (or spread across) the Hodge subspaces.
By optimizing dynamical systems formulated with Hodge Laplacian matrices, our work paves the way for future studies at the interface of  dynamical systems, combinatorial optimization, and algebraic topology.

%

The paper is organized as follows.  
In Sec.~\ref{sec:back}, we present background information on synchronization of the Kuramoto model, both on graphs and SCs, as well as the SAF optimization framework.  
In Sec.~\ref{sec:new tools}, we extend the SAF framework to HLK dynamics with projected dynamics onto lower- and upper-dimensional simplices within SCs.
In Sec.~\ref{sec:SAFoptimization}, we highlight the framework with computational experiments addressing several combinatorial optimization problems,  investigating the solutions' dependence on the Hodge subspaces and SC homology.
We present a conclusion in Sec.~\ref{sec:conclusion}.

\section{Background Information}\label{sec:back}

We provide background information on the following topics: In Sec.~\ref{sec:kuramoto}, we review the classic model of Kuramoto oscillators coupled over graphs. In Sec.~\ref{sec:SAFOnGraphs}, we present the SAF optimization framework. In Sec.~\ref{sec:back_simplicial}, we discuss SCs, boundary matrices and homology. In Sec.~\ref{sec:HodgeLaplacian}, we introduce Hodge Laplacian matrices and an associated (HLK) Kuramoto model. In Sec.~\ref{sec:phase locking}, we discuss conditions ensuring phase-locked synchronization for this model.

\subsection{Kuramoto Model for Graph-Coupled Phase Oscillators}\label{sec:kuramoto}
Kuramoto phase oscillators  \cite{kuramoto2003chemical} are a paradigmatic model for studying synchronization among heterogeneous dynamical systems. Notably, the model can be derived using first principles---a phase reduction approach \cite{strogatz2000kuramoto,nakao2016phase}  applied to a set of dynamical systems that are weakly coupled and each exhibiting a limit cycle oscillation.
The model is also widely celebrated in part due its analytical tractability for theory development (see, e.g., \cite{strogatz2000kuramoto,acebron2005kuramoto,ott2008low}) and has been used to study the effects for various structural and dynamical properties including graph-based network coupling \cite{rodrigues2016kuramoto,restrepo2005onset, restrepo2006synchronization}, time delays \cite{yeung1999time}, external forcing \cite{childs2008stability,antonsen2008external}, community structure \cite{skardal2012hierarchical},  higher-order sinusoidal coupling \cite{skardal2011cluster}, correlations \cite{coutinho2013kuramoto}, and adaptivity \cite{taylor2010spontaneous,skardal2014complex}.

We define the Kuramoto model by a group of $N$ oscillators, each having a natural frequency $\omega_i\in\mathbb{R}$ and phase $\theta_i\in[0,2\pi)$.  The phases evolve according to a system of sinusoidally coupled ordinary differential equations
\begin{equation} \label{eq:GraphKuramoto} 
    \dot{\theta_i} = \omega_i - \sigma \sum_{j=1}^N A_{ij} \sin{(\theta_i - \theta_j}),
\end{equation}
where scalar $\sigma\ge0$ is a  coupling strength that tunes the strength of interaction between pairs of oscillators. We define the vectors $\boldsymbol{\theta}=[\theta_1,\dots,\theta_N]$ and
$\boldsymbol{\omega}=[\omega_1,\dots,\omega_N]$.
The oscillators are coupled over a graph with an adjacency matrix $A$ in which the nonzero entries, $A_{ij}\ge0$, identify which pairs of oscillators are coupled.
%
 %
As is standard, 
we project the phases onto the unit circle to define
\begin{equation} \label{eq:order} 
r e^{i \psi(t)} = \frac{1}{N} \sum_{j=1}^N e^{i \theta_j (t)}.
\end{equation}
Scalar $r\ge0$ is known as the  Kuramoto order parameter, and it quantifies the extent of phase synchronization. We refer to angle $\psi(t)$, which can change with time, as the mean phase.
%
%
%

\begin{figure}[b!]
\centering
    \includegraphics[width=.7\linewidth]{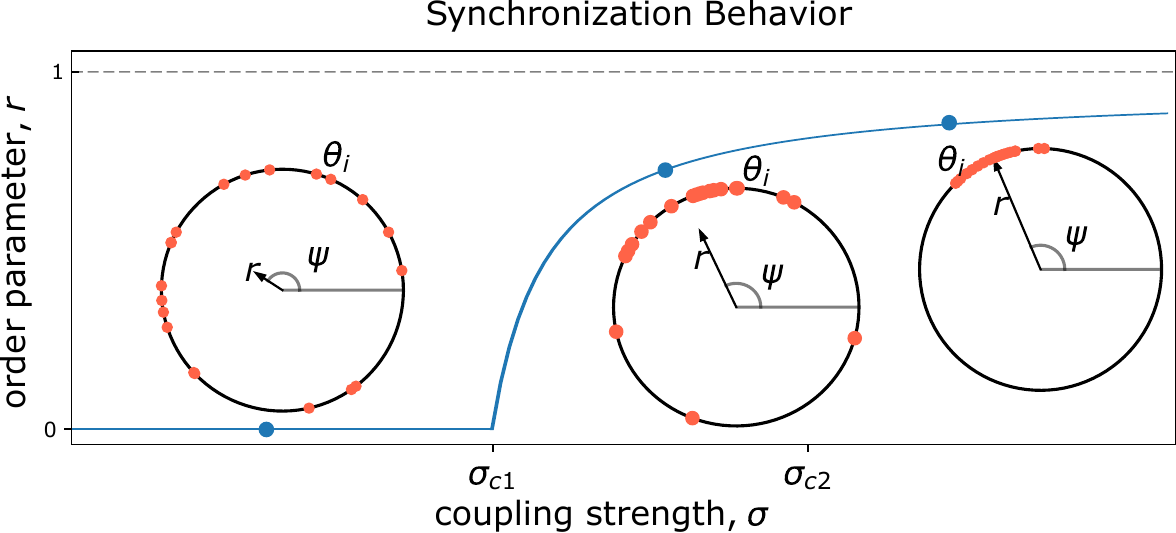}
    \vspace{-.3cm}
    \caption{{\bf Phase transitions for the Kuramoto  model.}  As one increases a coupling strength $\sigma$, Eq.~\eqref{eq:GraphKuramoto} exhibits three distinct phases:
    (i) $\sigma\in[0,\sigma_{c1}]$ yields an ``incoherent state'' in which no synchronization occurs as measured by $r\approx 0$;
    (ii) $\sigma\in(\sigma_{c1},\sigma_{c2}]$ yields a state with ``partial synchronization''  in which $r$ diverges from 0 due to some  phases clustering  near $\theta_i\approx \psi(t)$;  and
    (iii) $\sigma>\sigma_{c2}$ yields a ``phase locked'' state in which all oscillators approach fixed-point limits in the rotating frame $\theta_i\mapsto\theta_i-\psi(t)$. 
    %
    %
    %
    }
  \label{fig:Oscillators on Circle}
\end{figure}

In Figure \ref{fig:Oscillators on Circle}, we plot the Kuramoto order parameter $r$ for a system with varying $\sigma$, which reveals three distinct system phases. For small $\sigma$, synchronization doesn't occur and  $r \approx 0$. For intermediate $\sigma$, $r $ diverges from 0, indicating the onset of a synchronized cluster of phase oscillators (possibly with some oscillators that have yet to join the group and remain ``drifting''). For large $\sigma$, the system exhibits a ``phase locked'' state where all phases 
converge to limits within a  rotating reference frame: $\theta_i\mapsto\theta_i-\psi(t)$.
%
%
These phase transitions are central topics in the literature. The first phase transition point, $\sigma_{c1}$, is well-studied, e.g., for Kuramoto models with all-to-all coupling \cite{strogatz2000kuramoto} and network coupling \cite{restrepo2005onset,restrepo2006synchronization}.
 %
%
%
For network-coupled oscillators, the onset of global phase locking can be predicted by \cite{dorfler2013synchronization}
 \begin{equation}
    \label{eq:Synch Criteria for Networks}
    \sigma_{c2}=||L^\dagger \boldsymbol{\omega}||_{\infty, \mathcal{E}} ,
\end{equation} 
where $L^\dagger$ is the pseudo-inverse of the Laplacian matrix $L = D-A$, which uses a graph adjacency matrix $A$ and a diagonal matrix $D$ with entries $D_{ii}=\sum_j A_{ij}$ that encode the node degrees. Here, norm $||{\bf x}||_{\mathcal{E}, \infty} = \text{max}_{\{i, j\} \in \mathcal{E}} |x_i - x_j|$ to denote the largest pairwise dissimilarity across the graph's edges $\mathcal{E}=\{(i,j)|A_{ij}>0\}$.  The pseudo-inverse of the Laplacian matrix is significant for our approach for the optimization of synchronization, which extends the following framework.

\subsection{Synchrony Alignment Function (SAF) for Optimization \cite{skardal2014optimal}}\label{sec:SAFOnGraphs}
The SAF measures the ability for graph-coupled heterogeneous oscillators to synchronize. Because our main contribution is a generalization of the SAF, we briefly review its derivation and definition here.   Within the regime of strong synchronization ($r\approx 1$), Eq.~\eqref{eq:GraphKuramoto} can be linearized to yield
\begin{equation} \label{eq:GraphKuramotoLinearization}
    \frac{d\theta_i}{dt} = \omega_i - \sigma \sum_{j=1}^N L_{ij} \theta_j,
\end{equation}
or in vector form, 
$d\boldsymbol{\theta}/dt = \boldsymbol{\omega} - \sigma L\boldsymbol{\theta}$. 
For phase-locked oscillators,   $d \theta_i /dt = \Omega$ for each $i$, implying 
\begin{equation} \label{eq:GraphKuramotoPhaseLock}
   \Omega \mathbf{1} = \boldsymbol{\omega} - \sigma L \boldsymbol{\theta}^*.
\end{equation}
Here, $\mathbf{1}$ is a vector of ones and $\boldsymbol{\theta}^*$ is a vector of steady-state oscillator phases within a rotating reference frame.
Assuming the  graph is connected, the eigenvalues of a Laplacian $L$ can be ordered $0 = \lambda_1 < \lambda_2 \leq \lambda_3 \leq \dots \leq \lambda_N$, and we denote the associated eigenvectors by ${\bf v}^{(n)}$. In the case of an undirected graph (i.e., $A^T=A$) $L$ is symmetric and real, and therefore diagonalizable, and 
its Moore-Penrose inverse can be written $L^\dag =  \sum_{n=2}^N \lambda_i^{-1} \mathbf{v}^{(n)}\mathbf{v}^{(n)^T}$,
\cite{ben2003generalized} (with the term omitted for the zero-valued eigenvalue, $\lambda_1=0$). Notably, the multiplicity of the zero eigenvalue of the graph Laplacian is equal to the number of connected components (which is one for  a connected graph).

Rearranging Eq.~\eqref{eq:GraphKuramotoPhaseLock} and multiplying both sides by $\sigma^{-1} L^\dag$ yields 
\begin{equation}
    \label{eq:Angle Solution in Network}
    \boldsymbol{\theta}^* = \sigma^{-1}L^\dag \boldsymbol{\omega} - \sigma^{-1} L^\dag (\Omega \mathbf{1}) + {\bf v},
\end{equation}
where ${\bf v} \in \text{ker}(L)$ represents the part of $\boldsymbol{\omega}^*$ that is in the kernel of $L$.  Because $\text{ker}(L) = \text{ker}(L^\dagger) = \text{span}(\mathbf{1})$, it follows that ${\bf v} = c\mathbf{1}$ and the second term vanishes on the right-hand side of Eq.~\eqref{eq:Angle Solution in Network}.
Next, we multiply both sides of Eq.~\eqref{eq:Angle Solution in Network} by $N^{-1}\mathbf{1}^T$ to yield
\begin{equation}
    \overline{\theta} = \sigma^{-1}N^{-1}\mathbf{1}^T L^\dag \boldsymbol{\omega} + c,
\end{equation}
where $\overline{\theta}=N^{-1}\sum_i {\theta}_i^*$ is the arithmetic mean of the phases ${\theta}_i^*$.  Since $L$ is symmetric and $\mathbf{1} \in \text{ker}(L)$, $\mathbf{1}^T L^\dag = \mathbf{0}$, and so $c = \overline{\theta}$. Thus, Eq.~\eqref{eq:Angle Solution in Network} can be reduced to
\begin{equation}
    \label{eq:GraphSAFDerivation1}
    \mathbf{\theta^*} - \overline{\theta}\mathbf{1} = \sigma^{-1}L^{\dag}\boldsymbol{\omega} .
\end{equation}

We next consider the variance order parameter $R = 1-\sigma_\theta^2/2$, where $\sigma_\theta^2$ is the variance of the phases of the oscillators, which is a second-order approximation to the Kuramoto order parameter $r$ \cite{taylor2016synchronization}.  We have 
\begin{equation}
\label{eq:variance r}
\begin{aligned}
    R &= 1-\sigma_\theta^2/2    \\
    &=1-\frac{1}{2N}||\boldsymbol{\theta}^* - \overline{\theta}\mathbf{1}||^2  \\
    &= 1-\frac{1}{2N}||\sigma^{-1}L^\dag \boldsymbol{\omega}||^2 \\
    &= 1-J(\boldsymbol{\omega}, L)/2\sigma^2,
\end{aligned}
\end{equation}
where we have also defined the
%
Synchrony Alignment Function (SAF) \cite{taylor2016synchronization} for undirected networks  
\begin{equation} \label{eq:GraphSAF}
    J(\boldsymbol{\omega}, L)  = N^{-1} ||L^{\dag} \boldsymbol{\omega}||_2^2 = \frac{1}{N} \sum^N_{n=2} \frac{(\boldsymbol{\omega}^T {\bf v}^{(n)})^2}{\lambda_n^2}.
\end{equation}
Thus, when $R \approx 1$, we have that 
\begin{equation}
    \label{eq:SAF order parameter}
    r \approx 1-J(\boldsymbol{\omega}, L)/2\sigma^2.
\end{equation}  
The SAF $J(\boldsymbol{\omega}, L)$ can be computed for any frequency vector $\boldsymbol{\omega}$ and Laplacian matrix $L$, and small values indicate a network's propensity for strong synchronization. 
This allows for a study of networks to determine which structural--dynamical conditions promote optimal synchronization \cite{taylor2016synchronization,skardal2019synchronization,mikaberidze2025emergent}. 
While the SAF was originally developed for synchronization over undirected graphs, it has also been extended for directed graphs \cite{skardal2016optimal}, for systems with uncertainty about their structural and/or dynamical properties \cite{skardal2019synchronization}, and with hierarchical optimization for modular networks with both connected and disconnected configurations \cite{chamlagai2022grass}.

We also note that the SAF has  been extended to a different higher-order, Kuramoto model for synchronization that also uses SCs \cite{skardal2021higher}. While this is similar to the research that we present herein, there are several significant differences. First, similar to the original Kuramoto model, the model studied in \cite{skardal2021higher} represents each oscillator by a vertex/node (which is called a 0-dimensional simplex, or 0-simplex, in an SC). In contrast, the model we study represents each oscillator by a higher-dimensional simplex (e.g., a 1-simplex, which is analogous to an edge in a graph). 
Moreover, Hodge Laplacian matrices and their associated  mathematical tools from algebraic topology and homology are cornerstones for our present research; whereas these topics do not arise and are not relevant to the prior work \cite{skardal2021higher}. 
Thus, our work addresses an important gap in the literature.



\subsection{Simplicial Complexes (SCs), Boundary Matrices and Homology}\label{sec:back_simplicial}

Given a set $\mathcal{V}=\{1,\dots,N\}$ of vertices, graphs can encode dyadic connections between  pairs of two vertices. In contrast, a SC contains simplices that  more generally encode multi-way, polyadic interactions among sets of vertices. That is, a $k$-dimensional simplex (or simply, a $k$-simplex) $\mathcal{S}\subset \mathcal{V}$ is a set of $k+1$ nodes.  For example, a $0$-simplex is often called a \textit{node} or \textit{vertex},  a $1$-simplex is an \textit{edge}, a 2-simplex a ``filled in'' \textit{triangle}, a 3-simplex is a ``filled in'' \textit{tetrahedron}, and so on.

For a $k$-simplex $\mathcal{S}$, any $(k-1)$-simplex that is a subset of $\mathcal{S}$ with $k$ elements is called its  \textit{face}, while for any such face, $\mathcal{S}$ is called its \textit{coface}.  A \textit{simplicial complex} is a set of simplices in which any face of a simplex in the set must also be contained in the set, and where the intersection between faces is either another simplex in the set or it is the empty set.  For example, if a simplicial complex includes a triangle $\{a, b, c\}$, it must also contain the edges $\{a, b\}$, $\{b, c\}$, and $\{a, c\}$. This property sets simplicial complexes apart from hypergraphs, as the latter do not have the requirement of ``filling in'' hyperedges with all lower-dimensional hyperedges.  This difference allows the study of simplicial complexes as topological spaces.
The \textit{dimension} of a simplicial complex is the maximum dimension of its simplices.  An undirected graph can be interpreted as a 1-dimensional simplicial complex.  This paper primarily focuses on 2-dimensional simplicial complexes, although the results naturally extend to higher dimensions. 

\begin{figure}[t]
\centering
  \includegraphics[width=\linewidth]{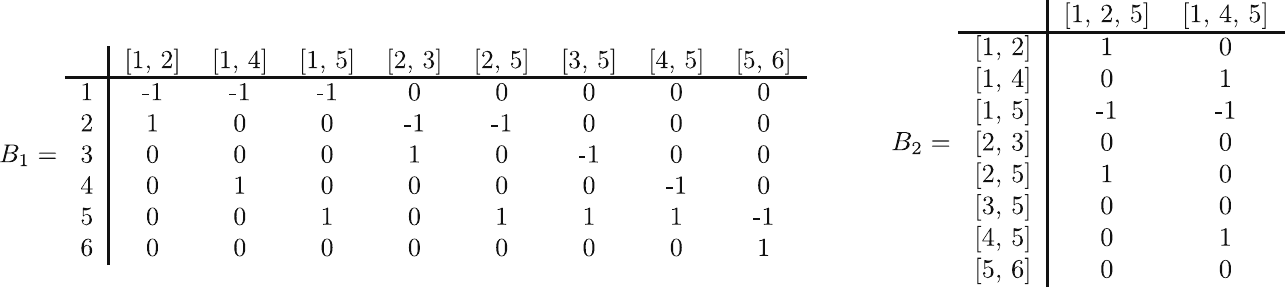}  
  \vspace{-.3cm}
  \caption{
  {\bf Boundary matrices  $B_1$ and $B_2$ for the SC shown in Fig.~\ref{fig:projections}}.
   As there are no simplices of dimension greater than $2$, this is a $2$-dimensional SC, and all boundary matrices beyond $B_2$ are empty.}
  \label{fig:boundary}
\end{figure}

For a simplicial complex to be represented as a topological space, each simplex must be oriented.  The orientation of a $k$-simplex is an equivalence class of the ordering of its nodes, where two classes are equivalent if an even number of pairwise permutations will convert one to the other. The straightforward method of orienting a simplicial complex begins by arbitrarily ordering the vertices $x_1, x_2, \dots x_n$.  Then the orientation of each simplex is simply its vertices in the order of increasing index.  The choice of ordering, as well as the method of determining an orientation for each simplex is necessary for topological reasons, but those choices do not need to signify a direction of the simplices, nor do they affect the dynamics that we study in this paper.
%
%
The \textit{boundary map} $\partial_k$ takes a $k$-simplicial complex to an alternating sum of its faces based on their orientation: 
\begin{equation}
    \label{eq:boundary operator}
    \partial_k([x_0, x_1, \dots, x_{k}]) = \sum_{i=0} (-1)^i [x_0, \dots, x_{i-1}, x_{i+1}, x_k]
\end{equation}
The matrix representation of this map is the \textit{boundary matrix} $B_k$, which encodes the connections between $k$-simplices with their $(k-1)-$simplex faces. An entry $[B_k]_{ij}$ of $B_k$ is $\pm1$ if the $i$th $(k-1)$-simplex is a face of the $j$th $k$-simplex (where the sign indicates orientation), and it is otherwise 0.  A simplicial complex is uniquely described by the collection of its boundary matrices.


In Fig.~\ref{fig:boundary}, as an example, we provide the boundary matrices for the simplicial complex shown in Fig.~\ref{fig:projections}.  
%
Observe that each column of $B_1$ corresponds to an edge $(i,j)$,  and all entries in a column are zero except for the $i$-th and $j$-th entries, which are $-1$ and $1$, respectively. The rows and columns of $B_2$  correspond to edges and 2-simplices, and values of $\pm 1$ indicate incidences between faces and cofaces. The signs indicate whether or not the edges' orientations match that for a triangular cycle/boundary around each 2-simplex.  For example, the orientation of $(1,5)$ does not match that of the 2-simplex $(1, 2, 5)$.

The boundary functions for an SC allow one to  study its homology, which describes the presence of ``holes'' in the SC. We  define the $n$-th (simplicial) homology group $H_n$ as $\text{Ker}(\partial n)/\text{Im}(\partial n+1)$. The dimension of  $H_n$ is called the $n$-th Betti number,  the zeroth Betti number  is  equal to the number of connected components in a SC. The 1st Betti number defines the number of 1-dimensional cycles that are not “filled-in” by triangles (i.e., they are cyclic paths other than the boundaries of  2-simplices).  A cycle can be thought of as a 1-dimensional ``hole'' (with a 1-dimensional boundary), and one can also study higher-dimensional "holes" which are counted by higher-dimensional Betti numbers.

In Fig.~\ref{fig:projections}, also observe that there are three triangles giving rise to three different length-3 paths that are cycles. Two of these cycles are boundaries of 2-simplices. For example, the 2-simplex $(1, 2, 5)$ has a boundary consisting of three 1-simplices, $\{(1,2)$, $(2,5)$, $(5,1) \}$. 
However the length-3 path $\{(2,3)$, $(3,5)$, $(5,2) \}$ is not the boundary of 2-simplex, but it instead  identifies a 1-dimensional hole, or 1-cycle, in the SC. Thus the SC has a  1-dimensional  homology $H_1$, and the first Betti number is 1 for this example. (The zeroth Betti number is 0, since the SC does not have disconnected components.)
%
Later, we will study how homology influences optimal synchronization over SCs.

\subsection{Hodge Laplacian Kuramoto (HLK) Model}
\label{sec:HodgeLaplacian}

For a given dimension $k$, the Hodge Laplacian matrix is defined as 
\begin{equation}
    \label{eq:HodgeLaplacian}
    L_k = B_k^TB_k + B_{k+1}B_{k+1}^T,
\end{equation}
which is a size-$n_k$ matrix in which each entry $[L_k]_{ij}$  describes the interaction between the $i$-th and $j$-th $k$-simplices. Two $k$-simplices are called \textit{lower adjacent} if they share a $(k-1)$-dimensional face, and they are called \textit{upper adjacent} if they share a $(k+1)$-dimensional coface, and these two interaction types are represented, respectively, by the two terms in the right-hand-side of Eq.~\eqref{eq:HodgeLaplacian}.    For example, two edges are lower adjacent if they share a common node, and the edges are upper adjacent if they are both on the boundary of a 2-simplex. 
To help study  these interactions separately, we   adopt the notation  $L_k^{[-]} = B_k^T B_k$ and $L_k^{[+]} = B_{k+1}B^T_{k+1}$ so that $L_k = L_k^{[-]}+L_k^{[+]}$. Note that any nonzero eigenvalue of the Hodge Laplacian matrix must also be a nonzero eigenvalue of either $L_k^{[-]}$ or   $L_k^{[+]}$ (i.e., this occurs because all of their eigenvalues are nonnegative). Moreover, the eigenvectors associated with positive eigenvalues of $L_k^{[-]}$ are orthogonal to those associated with the nonzero eigenvalues of $L_k^{[+]}$. Thus one can consider the null space of matrix $L_k$ and the image spaces of $L_k^{[-]}$ and $L_k^{[+]}$ (or equivalently those of  $\text{im}(B_k^T)$ and $\text{im}(B_{k+1})$) to partition the vector space $\mathbb{R}^{n_k}$ into three orthogonal subspaces,
which is known as the 
%
Hodge Decomposition 
\begin{equation}
    \label{eq:HodgeDecomp}
    \mathbb{R}^{n_k} = \text{im}(B_k^T) \oplus\text{im}(B_{k+1})\oplus\text{ker}(L_k).
\end{equation}
When $k=1$, $L_k$ describes connections between 1-simplices (i.e., edges) through either their adjacent 0-simplices (i.e., nodes) or adjacent 2-simplices (i.e., filled triangles).  In this case, the Hodge subspaces are often referred to as the gradient, curl, and harmonic subspaces. Notably, the dimension of the harmonic subspace  equals the multiplicity of the zero-valued eigenvalue of $L_k$, and it is also equal to the number of 1-cycles in the SC, i.e., the first Betti number  \cite{barbarossa2020topological}.  We will make use of these spectral properties for optimization problems and solutions presented later in Sec.~\ref{sec:SAFoptimization}.

The Kuramoto model can be extended to SCs using boundary matrices.  First, we note that the original model (on graphs) can also be written as $\dot{\theta} = \omega - \sigma B_1 \sin{(B_1^T \theta)}$. Thus motivated, we consider the Hodge Laplacian Kuramoto (HLK) dynamics   \cite{millanexplosive}
\begin{equation}\label{eq:HLK Dynamics}
    \dot{\boldsymbol{\theta}} = \boldsymbol{\omega} - \sigma_{[-]} B^T_k \sin{(B_k \boldsymbol{\theta}) - \sigma_{[+]} B_{k+1} \sin{(B^T_{k+1} \boldsymbol{\theta})}}.
\end{equation}
Note that the graph-coupled Kuramoto model is recovered with $k = 0$, since by construction, $B_0 = 0$.
%
We note that the original formulation of HLK dynamics assumed  $\sigma_{[\pm]}  =\sigma$ \cite{millanexplosive}, and the form  presented above   stems from \cite{nurisso2024unified},  where $\sigma_{[\pm]}$ represent distinct coupling strengths for the lower- and upper-dimensional interactions.  
In Sec.~\ref{section:balancing}, we will reparametrize $\sigma_{[\pm]}$ further to facilitate optimization while conserving a measure of total coupling strength.

We further highlight that the model defined by Eq.~\eqref{eq:HLK Dynamics} assigns phases $\theta_i$ and frequencies $\omega_i$ to $k$-simplices in an SC, and we can use boundary matrices to project these down to the $(k-1)$-simplices and up to the $(k+1)$-simplices. That is, we define
    $\boldsymbol{\theta}^{[-]} = B_k \boldsymbol{\theta},$ $\boldsymbol{\omega}^{[-]} = B_k \boldsymbol{\omega}$,
    $\boldsymbol{\theta}^{[+]} = B^T_{k+1} \boldsymbol{\theta}$, and $\boldsymbol{\omega}^{[+]} = B^T_{k+1} \boldsymbol{\omega}$ so that ${\theta}^{[-]}_i$ and ${\omega}^{[-]}_i$ assign a phase and frequency to the $i$-th simplex with dimension $k-1$  and ${\theta}^{[+]}_i$ and ${\omega}^{[+]}_i$ assign a phase and frequency to the $i$-th simplex with dimension $k+1$.
%
Using the Hodge Decomposition, it has been shown  that these projections lead to associated, simultaneous dynamics on the $(k-1)$-simplices and  $(k+1)$ simplices \cite{millanexplosive} :
\begin{align}\label{eq:KuramotoDecoupledDown}
    \dot{\boldsymbol{\theta}}^{[-]} &=  \boldsymbol{\omega}^{[-]}  - \sigma_{[-]} L^{[+]}_{k-1} \sin{(\boldsymbol{\theta}^{[-]})}\\
    \label{eq:KuramotoDecoupledUp}
    \dot{\boldsymbol{\theta}}^{[+]} &= \boldsymbol{\omega}^{[+]} - \sigma_{[+]} L^{[-]}_{k+1} \sin{(\boldsymbol{\theta}^{[+]})}.
\end{align} 
That is, the dynamics from Eq.~\eqref{eq:HLK Dynamics} effectively yields two dynamical systems defined over simplices at two different dimensions.  In the case   $k=1$, the (unprojected)  HLK oscillators are originally assigned to 1-simplices (i.e., edges), whereas the projections respectively yield two associated dynamical systems with oscillators assigned to
0-simplices and 2-simplices (i.e., nodes and triangles).  
%

\begin{figure*}[t!]
    \includegraphics[width=\linewidth]{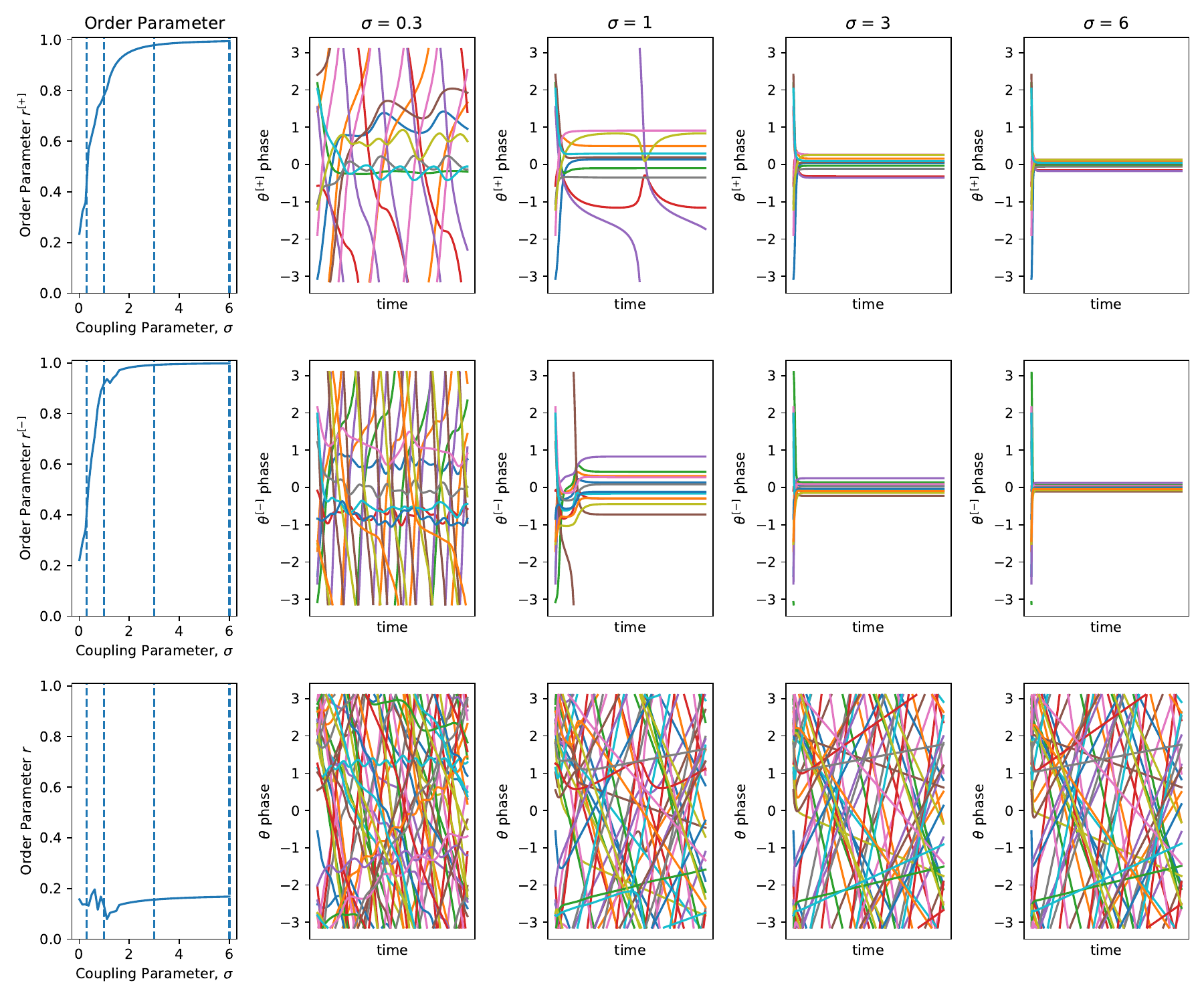}
    \label{fig:SynchSimulations}
    \vspace{-.5cm}
  \caption{{\bf HLK dynamics for projected  and  unprojected dynamics.}   
  In the leftmost column, we show Kuramoto order parameters versus  coupling strength $\sigma_{[\pm]}=\sigma$ for the upper- and lower-projected dynamics (top and middle rows, respectively) as well as the unprojected dynamics (bottom row). Vertical dashed lines indicate $\sigma\in\{0.3,1,3,6\}$ values for which the remaining columns depict time series for the  oscillators' phases   (i.e., $\boldsymbol{\theta}^{[\pm]}$ and $\boldsymbol{\theta}$).  Observe that synchronization occurs for the two projected systems for sufficiently large $\sigma$, but not for the  unprojected dynamics.
  }
\end{figure*}

Synchronization for the two projected dynamics can be quantified using two separate order parameters
\begin{align}
\label{eq:ProjectedOrderParameters}
    r^{[\pm]}  &= \frac{1}{N^{[\pm]}} \left | \sum_{j=1}^{N^{[\pm]}} e^{i \boldsymbol{\theta}^{[\pm]}_j}\right | ,
\end{align}
where $N^{\pm}$ is the number of $(k\pm 1)$-simplices.  
 %
In Fig.~\ref{fig:SynchSimulations}, we show Kuramoto order parameters versus $\sigma$ (leftmost column) and time series for the oscillator phases 
for the (top row) upper-projected dynamics, (middle row) lower-projected dynamics, and (bottom row) unprojected dynamics. 
We simulate  HLK dynamics on a SC containing of twelve 0-simplices, thirty 1-simplices, and nineteen 2-simplices for different $\sigma$. See Appendix~\ref{appendix:small SC} for a visualization and description of the SC. Observe that synchronization and phase locking occurs for the lower- and upper-projected systems when $\sigma$ is sufficiently large, but it does not occur for the unprojected system. In the next section, we discuss conditions that allow synchronization and guarantee phase locking for these respective phase-oscillator systems.

Before continuing, we highlight a significant difference between the original graph-coupled Kuramoto model [Eq.~\eqref{eq:GraphKuramoto}] and the HLK model [Eq.~\eqref{eq:HLK Dynamics}] that will be important later for our optimization analysis. Specifically, Eq.~\eqref{eq:GraphKuramoto} is often paired with a mod-$2\pi$ operation so that the oscillator phases remain within $\theta_i\in[0,2\pi)$ (although it is sometimes convenient to define $\theta_i\in[-\pi,\pi)$). The same mod-$2\pi$ pairing can be combined with Eq.~\eqref{eq:HLK Dynamics}; however, this is not true for the projected systems [Eq.~\eqref{eq:KuramotoDecoupledDown} and \eqref{eq:KuramotoDecoupledUp}], since the projection operators (i.e., multiplication by $B_k$ or $B_{k+1}^T$) do not commute with the mod-$2\pi$ operator. It follows that the projected phases $ \boldsymbol{\theta}^{[\pm]}$ do not necessarily remain within the range $[0,2\pi)$, and this will later  complicate our linearization of the dynamics near $\boldsymbol{\theta}^{[\pm]} \approx0$ in Sec.~\ref{sec:HLK SAF}.

\subsection{Synchronization Conditions for HLK Dynamics}
\label{sec:phase locking}
In this section, we review  synchronization properties that arise from the model introduced in Sec.~\ref{sec:HodgeLaplacian}, starting with the projected phases: $\boldsymbol{\theta}^{[-]}$ and $\boldsymbol{\theta}^{[+]}$. 
As shown in \cite{nurisso2024unified}, for any SC there exists a sufficiently large $\sigma$ that ensures  both the lower and upper dimensional projections phase lock.  In this state, $\dot{\boldsymbol{\theta}}^{[-]} \to \mathbf{0}$ and $\dot{\boldsymbol{\theta}}^{[+]} \to \mathbf{0}$, so the projected phases no longer change with time.  Furthermore, as the coupling strength increases, both systems have phase-locked oscillators and their phases can converge to $0$ (mod $2 \pi$).  Thus, in the limit of infinite coupling strength, we obtain full phase synchronization whereby all  projected phases approach $0$ (mod $2 \pi)$.  Notably, the required coupling strength for phase locking will be different for the two projections
\cite{nurisso2024unified}
\begin{align} \label{eq:sigma fp}
    \sigma_{fp}^- &= ||(B_k^T)^\dag \boldsymbol{\omega}||_2 \\
    \sigma_{fp}^+ &= ||(B_{k+1})^\dag \boldsymbol{\omega}||_2.
\end{align}
That is, if $\sigma_{[\pm]} \geq \sigma_{fp}^{\pm}$, then the lower/upper projection is guaranteed to phase lock after a finite amount of time.  
Note that these criteria represent sufficient, but not necessary, conditions to guarantee phase locking (and phase locking can potentially occur more generally).
Synchronization of the unprojected system is less straightforward, and it can be studied by examining the phase component for each of the three Hodge subspaces.  The phase component in the gradient and curl subspaces both phase lock in the same manner as the projected phases \cite{nurisso2024unified}, since they are influenced, respectively, by the lower and upper components of the Hodge Laplacian. 
That is, as long as phase locking occurs for the projected dynamics, both the gradient and curl components of the unprojected  phases will eventually become constant.  
The harmonic component, on the other hand, is unaffected by the HLK dynamics since it lives in the null space of a Hodge Laplacian. Therefore, the harmonic component of oscillator phases, $\boldsymbol{\theta}^{(h)}$, will be continually driven by the harmonic component of the frequency vector, $\boldsymbol{\omega}^{(h)}$. 
Thus, 
the unprojected phases will continue to oscillate and synchronization is not possible for the unprojected system unless $\boldsymbol{\omega}^{(h)}= \mathbf{0}$. This can occur if
the harmonic subspace is empty (i.e., the 1-dimensional homology is empty) or if the harmonic subspace is non-empty but the  original frequency vector $\boldsymbol{\omega}$ is orthogonal to the harmonic subspace. As such, one should generally expect that synchronization will commonly occur for the projected HLK dynamics, but it can be rare for the unprojected (i.e., original) HLK dynamics.

%



\section{Optimizing Synchronization for the Projected HLK Dynamics} \label{sec:new tools}
We now present a theoretical framework to optimize phase synchronization under the HLK model. In Sec. \ref{section:balancing}, we introduce a balancing parameter to the HLK model that can tune the relative onset of synchronization for the two projections.
In Secs.~\ref{sec:HLK SAF} and \ref{sec:Upper SAF}, we extend the SAF framework to the lower- and upper-projected HLK dynamics, respectively. 

\subsection{Balancing the Lower and Upper Dimensional Interactions}
\label{section:balancing}

We reparameterize the HLK model by introducing a balancing parameter $\delta \in [-1,1]$ to strategically tune the  relative influence of lower- and upper-dimensional interactions,
   $ \sigma_{[\pm]} = \sigma (1\pm \delta),$
which yields the reformulated HLK dynamics   
\begin{equation}
    \label{eq:Generalized SC Kuramoto}
     \dot{\boldsymbol{\theta}} = \boldsymbol{\omega} - \sigma (1 - \delta) B^T_k \sin{(B_k \boldsymbol{\theta})} - \sigma (1 + \delta) B_{k+1} \sin{(B^T_{k+1} \boldsymbol{\theta})} .
\end{equation}
Note that the sum $\sigma_{[+]}+\sigma_{[-]}$ is always $2\sigma$, thereby conserving  a measure for the ``total'' coupling strength regardless of $\delta$.  We also highlight that a simple linearization of Eq.~\eqref{eq:Generalized SC Kuramoto} yields 
$ \dot{\boldsymbol{\theta}} = \boldsymbol{\omega} - \sigma L^{(-\delta)}_k\boldsymbol{\theta} $, where 
$L^{(\delta)}_k = (1 + \delta) B^T_k B_k +  (1 - \delta) B_{k+1}  B^T_{k+1} $  is a balanced Hodge Laplacian that has been introduced to study consensus dynamics over SCs  \cite{ziegler2022balanced}.

\begin{figure}[h!]
    \centering
    \includegraphics[width=\linewidth]{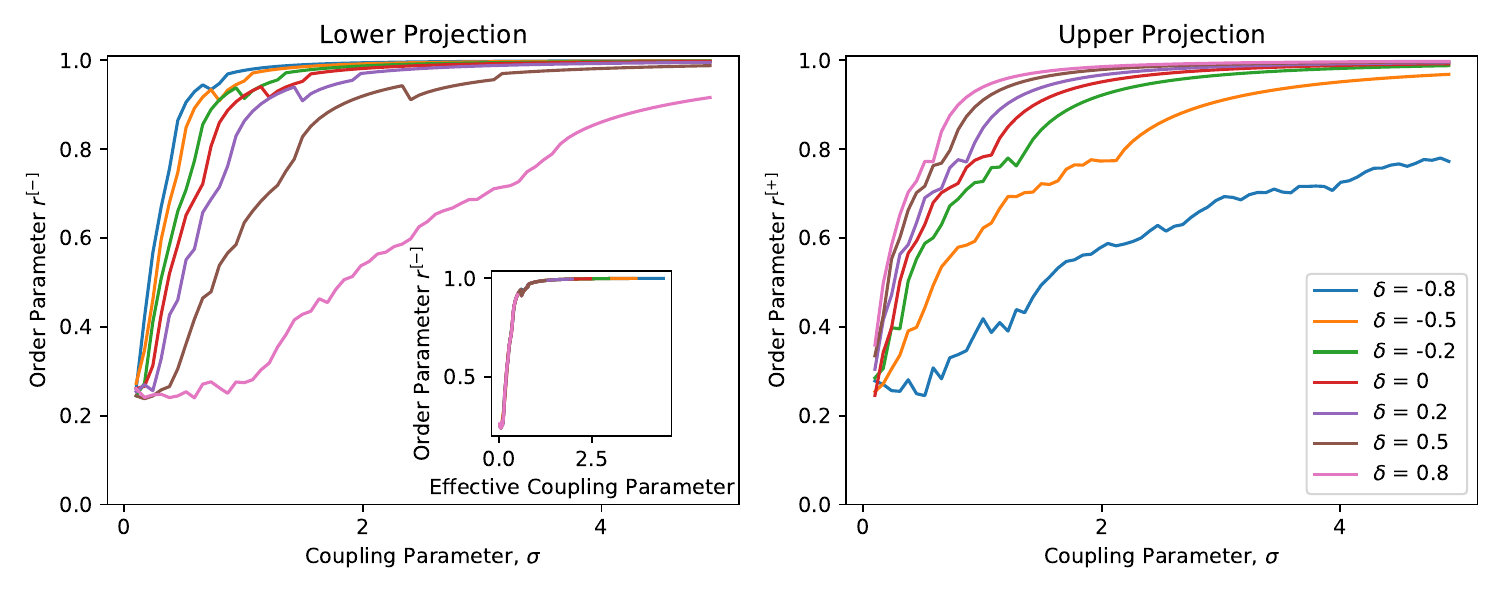}
    \label{fig:Delta Order Parameter}
    \vspace{-.5cm}
    \caption{{\bf Balancing the projected HLK dynamics.} We plot order parameters $ r^{[\pm]}$ versus $\sigma$ for the lower- and upper-projected HLK dynamics for choices of  the balancing parameter $\delta$.
    Varying $\delta$ tunes the relative onset of synchronization for these two projected systems, and the inset shows that all curves effectively collapse onto a single curve if one plots the order parameters versus $\sigma_{[\pm]}$ rather than $\sigma$. 
    }
\end{figure}

In Fig.~\ref{fig:Delta Order Parameter}, we highlight that the introduction of $\delta$ allows us to tune the relative onset of  synchronization  for the lower- and  upper-dimensional projections.
%
For various choices of $\delta$, we plot the order parameters $ r^{[\pm]}$ versus $\sigma$, and the $r^{[\pm]}$ values are numerically calculated as a time average following a long simulation duration that ensures  transient dynamics have decayed and the phases are very near their phase-locked values (if it occurs). 
%
As expected, the projected systems synchronization for sufficiently large $\sigma$.  Observe that $r^{[+]}$  increases (and $r^{[-]}$ to decreases) for large positive $\delta$, and the opposite occurs for negative $\delta$.  
The inset figure  shows that the curves all collapse onto a single curve if the x-axis is rescaled by $(1 \pm\delta)$.  Thus,  the balancing parameter $\delta$ does not change the  curve shape of these transitions, but rather it introduces a dilation that controls which projection synchronizes first.

Our later development of optimization theory will focus on the scenario in which  phase locking occurs for both projected systems. 
To theoretically ensure phase locking, we expand on the phase-locking criteria $\sigma_{fp}^{\pm}$ given by Eq.~\eqref{eq:sigma fp} in Sec.~\ref{sec:phase locking}.  A straightforward application of that previous result implies that if $\sigma(1\pm \delta) \geq \sigma_{fp}^{\pm}$, then the lower/upper projection is guaranteed to phase lock after a finite amount of time.  To ensure that both  criteria are simultaneously met (i.e.,  so both projections synchronize), we derive the criterion 
\begin{equation}
    \label{eq:allowable delta}
    \frac{\sigma_{fp}^+}{\sigma}-1 \leq \delta \leq 1-\frac{\sigma_{fp}^-}{\sigma}.
\end{equation}
As $\delta \in [-1,1]$, we also have that for the lower/upper projection individually to be guaranteed to reach phase locking for some value of $\delta$, it is required that 
\begin{equation}
    \sigma \geq \sigma_{fp}^{\pm}/2.
\end{equation}
For both projections to meet this condition, we also need $\frac{\sigma_{fp}^+}{\sigma}-1 \leq  1-\frac{\sigma_{fp}^-}{\sigma}$ so that such a $\delta$ can exist in between them.  This is equivalent to 
\begin{equation}
    \sigma \geq \frac{1}{2}(\sigma_{fp}^+ + \sigma_{fp}^-) .
\end{equation}

\begin{figure*}[h!]
    \centering
    \includegraphics[width=.9\linewidth]{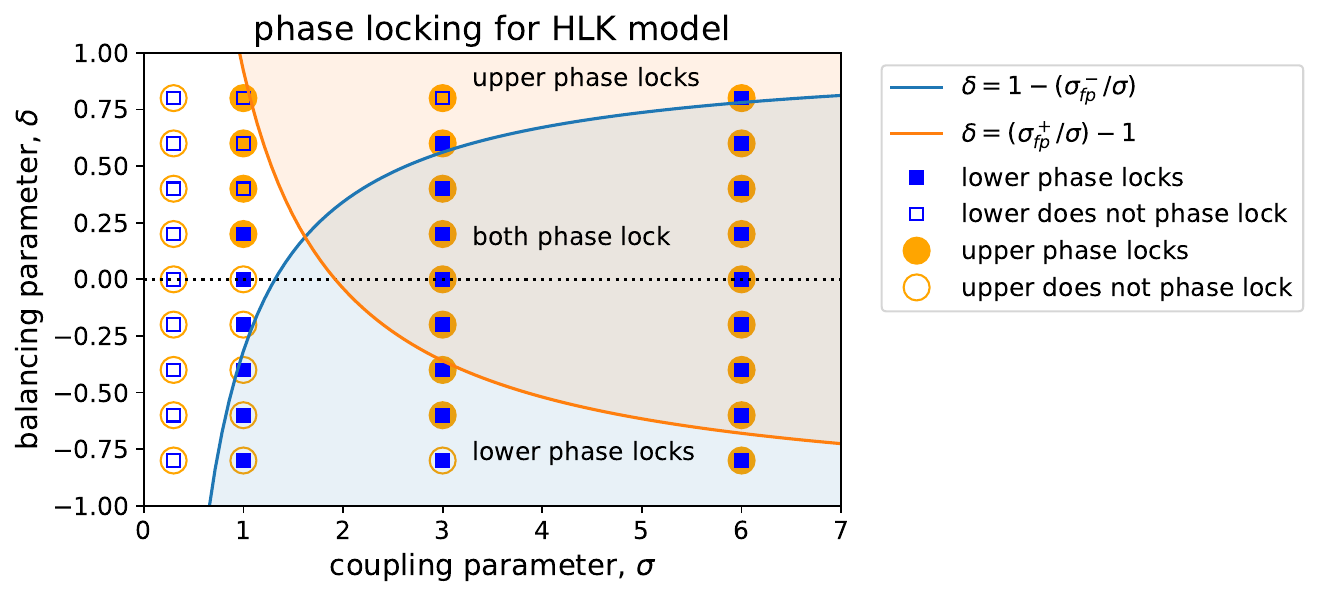}
    \label{fig:deltaCurvesSynch}
    \vspace{-.1cm}
  \caption{{\bf Criteria for phase locking across the $(\delta,\sigma)$ parameter space}. We compare theoretical predictions (curves and shaded regions) to numerical observations (symbols) of phase locking for the lower- and upper-projected HLK dynamics for different $\delta$ and $\omega$.
  Simulations were implemented with the same SC and parameters as in Figs.~\ref{fig:SynchSimulations}--\ref{fig:Delta Order Parameter}.
  Curves and shaded regions indicate the values of $\delta$ that guarantee phase locking in the lower-projection (blue) and  upper-projection (orange) of the dynamics.  
  We find  that both projected dynamics phase lock within the predicted regions (which are sufficient but not necessary conditions).
  }
\end{figure*}

In Fig.~\ref{fig:deltaCurvesSynch}, we validate these phase locking criteria by comparing theoretically predicted (shaded regions) and numerically observed (symbols) phase-locking behavior for the simulation results previously shown in Figs.~\ref{fig:SynchSimulations} and \ref{fig:Delta Order Parameter}. As expected, the above criteria are sufficient (but not necessary) to guarantee that phase locking  occurs for each projected system.

\subsection{SAF Extension to the Lower-Projected HLK Model}
\label{sec:HLK SAF}

In this section, we study the lower-dimensional projection of HLK dynamics given by Eq.~\eqref{eq:KuramotoDecoupledDown} and  develop a theoretical framework to maximize the order parameter  $ r^{[-]}$  given by Eq.~\eqref{eq:ProjectedOrderParameters}. To this end, we  extend the SAF optimization framework introduced in  Sec.~\ref{sec:SAFOnGraphs}.
%
%
For simplicity, we will restrict our scope to a 2-dimensional SC in which the oscillators are assigned to 1-simplices (i.e., the edges).  That is, we   study Eq.~\eqref{eq:KuramotoDecoupledDown} with $k=1$.
%
%

Similar to the derivation of the original SAF,
 we will develop theory for the
regime of strong phase synchronization, $ r^{[-]}\approx1$, and
assume that the oscillators are in the phase-locked state with
$\dot{\boldsymbol{\theta}}^{[-]} = \mathbf{0}$ and 
%
reach limiting values $\boldsymbol{\theta}^{[-],*}$.
It follows that Eq.~\eqref{eq:KuramotoDecoupledDown} reduces to 
\begin{equation}
    \sigma_{[-]} L_0^{[+]} \sin{(\boldsymbol{\theta}^{[-],*})}=\boldsymbol{\omega}^{[-]}.
\end{equation}
Because  $(-1)$-dimensional simplices do not exist, $L_0^{[+]} = L_0$, which is the normally defined graph Laplacian matrix.  Recall for the earlier SAF derivation in Sec.~\ref{sec:SAFOnGraphs} that a first step involved linearizing the sinusoidal term to obtain $L_0$. In contrast, for our  current derivation, $L_0$ arises  prior to linearization, and  we can therefore delay linearization to a later step (and future studies may even choose not to linearize).
Using that $\boldsymbol{\omega}^{[-]}=B_1 \boldsymbol{\omega} \in \text{Im}(B_1) = \text{Im}(L_0)$, we 
multiply both sides by the  Moore-Penrose pseudoinverse $\sigma_{[-]}^{-1}L_0^{\dag}$
\cite{wang2018generalized} and simplify to find
\begin{equation}
    \label{eq:DownSAF1}
    \sin(\boldsymbol{\theta}^{[-],*}) = \sigma_{[-]}^{-1}L_0^\dag \boldsymbol{\omega}^{[-]} + \mathbf{v}.
\end{equation}
Here, vector $\mathbf{v} \in \text{ker}(L_0^\dag) = \text{ker}(L_0)$ represents the part of $ \sin(\boldsymbol{\theta}^{[-],*})$ in the kernel of $L_0$, and as long as the SC is connected, this is given by $\text{ker}(L_0)=\text{span}(\mathbf{1})$.  

At this point, we would like to linearize the left-hand side, but 
it is not necessarily true that $\boldsymbol{\theta}^{[-],*} \approx \mathbf{0}$.  Instead, it must be the case that $\boldsymbol{\theta}^{[-],*}  \approx \mathbf{0}\ (\text{mod} \ 2\pi)$. (Recall our discussion at the end of Sec.~\ref{sec:HodgeLaplacian} about how the mod-$2\pi$ operator does not commute with the projection operators.)   Noting that $\boldsymbol{\theta}^{[-]} =B_1 \boldsymbol{\theta}$, we highlight that   a simple reduction of $\boldsymbol{\theta}^{[-],*}  (\text{mod} \ 2 \pi)$ may lead to an ``unreachable'' value that is outside 
the image space of matrix $B_1$. To remedy this, 
we re-parameterize to define the vector $\boldsymbol{\theta}^{[-],*}-2\pi\mathbf{z}$, where $\mathbf{z} \in \mathbb{Z}^{N^{[-]}}$ is the vector of integers such that  all components of $\boldsymbol{\theta}^{[-],*}-2\pi\mathbf{z}$ are in $[-\pi, \pi)$.  It follows that we can linearize by
\begin{equation}
    \sin(\boldsymbol{\theta}^{[-],*}) =  \sin(\boldsymbol{\theta}^{[-],*}-2\pi\mathbf{z}) \approx \boldsymbol{\theta}^{[-],*}-2\pi\mathbf{z},
\end{equation}
which yields
\begin{equation}
    \label{eq:DownSAF11}
    \boldsymbol{\theta}^{[-],*}-2\pi\mathbf{z} \approx \sigma_{[-]}^{-1}L_0^\dag \boldsymbol{\omega}^{[-]} + \mathbf{v}.
\end{equation}
The introduction of ${\bf z}$ was not required for the original SAF. Next, we let
$\mathbf{v} = c\mathbf{1}$ for some constant $c$, which we solve for by left-multiplying both sides by $\mathbf{1}^T/N^{[-]}$; 
\begin{equation}
    \label{eq:Down SAF Deriv Part}
    \overline{\boldsymbol{\theta}^{[-],*}} -2\pi \overline{\mathbf{z}} = \frac{\mathbf{1}^T L_0^\dag  \boldsymbol{\omega}^{[-]}}{N^{[-]} \sigma_{[-]}} + c.
\end{equation}
Here, the bars represent the arithmetic mean.  Next, we again use the fact that $L_0$ is symmetric and $\ker(L_0) =\ker(L_0^T)= \text{span}(\mathbf{1})$ to find $\mathbf{1}^T L_0^\dag =0$.  It also follows that $ \overline{\boldsymbol{\theta}^{[-],*}} =\mathbf{1}^T{\boldsymbol{\theta}^{[-],*}}/N^{[-]}=\mathbf{1}^TB_1\boldsymbol{\theta}^{*}/N^{[-]}=0$, since each column of $B_1$ contains zeros except for two entries (which are $1$ and $-1$), and so left-multiplication by $\mathbf{1}$ yields zero.
%
%
Thus, Eq.~\eqref{eq:Down SAF Deriv Part}  reduces to $c=-2\pi \overline{\mathbf{z}}$, implying
\begin{equation}
    \boldsymbol{\theta}^{[-],*}  = \sigma_{[-]}^{-1}L_0^\dag \boldsymbol{\omega}^{[-]} + 2\pi(\mathbf{z}-\overline{\mathbf{z}}\mathbf{1}).
\end{equation}

To proceed, we change the reference frame $\boldsymbol{\theta}^{[-]}_i \rightarrow \boldsymbol{\theta}^{[-]}_i-2\pi(\mathbf{z}_i-\overline{\mathbf{z}})$, which does not modify $\overline{\boldsymbol{\theta}^{[-],*}}$ (or the order parameter $r^{[-]}$, to first order).  In the new reference frame, the phase-locked states take the form   
\begin{equation}
    \boldsymbol{\theta}^{[-],*} = \overline{\boldsymbol{\theta}^{[-],*}}\mathbf{1} + \sigma_{[-]}^{-1}L_0^\dag  \boldsymbol{\omega}^{[-]}.
\end{equation}
Recalling that one can linearly approximate the Kuramoto order parameter with a variance order parameter (Sec.~\ref{sec:SAFOnGraphs}), we define
\begin{align}
R^{[-]} &= 1-\sigma_{\theta^{[-],*}}^2/2 \\
    &= 1-\frac{1}{2N^{[-]}}||{\boldsymbol{\theta}^{[-],*}} - \overline{\boldsymbol{\theta}^{[-],*}} ||^2 \\
    &= 1-\frac{1}{2N^{[-]}}||\sigma_{[-]}^{-1}L_0^{\dag} \boldsymbol{\omega} ||^2.
\end{align}
Thus, when $r^{[-]} \approx 1$, we have that 
\begin{equation}
    \label{eq:Down SAF order parameter}
    r^{[-]} \approx 1-J^{[-]}(\boldsymbol{\omega}^{[-]},  L_0)/2\sigma_{[-]}^2,
\end{equation} 
and we define the lower-projection SAF 
\begin{equation} \label{eq:LowerSAF}
    J^{[-]}(\boldsymbol{\omega}^{[-]}, L_0)  = \frac{ ||L_0^{\dag} \boldsymbol{\omega}^{[-]}||_2^2}{N^{[-]}} = \frac{1}{N^{[-]}} \sum_{n:\lambda_n\not=0}^{N^{[-]}} \frac{[(\boldsymbol{\omega}^{[-]})^T {\bf v}_{[-]}^{(n)}]^2}{\lambda_n^{[-]2}},
\end{equation}
where  ${\bf v}_{[-]}^{(n)}, \lambda_n$ are defined as the eigenvector/eigenvalue pairs for $L_0$.  Note that $ J^{[-]}(\boldsymbol{\omega}^{[-]}, L_0)$ can be interpreted as an objective function that can be strategically optimized to enhance synchronization for the lower-projected HLK model. For example, synchronization for the lower-projected dynamics will be enhanced if one aligns the projected frequencies $\boldsymbol{\omega}^{[-]}$ with eigenvectors associated with large eigenvalues $\lambda_n$ of $L_0$.

Before continuing, we note that functionally, the lower-projection SAF, $J^{[-]}$, is identical to the original SAF, $J$, given by Eq.~\eqref{eq:GraphSAF}. The only difference here is that we now denote the number of 0-simplices  (i.e., vertices) by $N^{[-]}$ and the first argument is the projected frequencies $\boldsymbol{\omega}^{[-]}=B_1 \boldsymbol{\omega}$. Despite this equivalence, we will use the `$-$' notation to help distinguish results for the lower- and upper-dimensional projections.

\subsection{SAF Extension to the Upper-Projected HLK Model}\label{sec:Upper SAF}
We now extend the SAF optimization framework to the upper-projected HLK model given by Eq.\eqref{eq:KuramotoDecoupledUp}, seeking to maximize the order parameter $r^{[+]}$ given by Eq.~\eqref{eq:ProjectedOrderParameters}. 
In the previous section, we found that the SAF is slightly more difficult to derive for the lower-projected HLK model  due to the projection operator $B_1$ (which does not commute with the mod-$2\pi$ operator). Nevertheless, both derivations yielded an identical functional form   (i.e., compare Eq.~\eqref{eq:LowerSAF} to Eq.~\eqref{eq:GraphSAF}).
%
For the upper projection, the derivation is again made difficult because of a projection operator (i.e.,  multiplication by $B_2^T$). Moreover, in this case one must consider a matrix  $L_2^{[-]}$ (i.e., as opposed to $L_0$). Unfortunately, its kernel cannot be expressed in a simple way, which further complicates the SAF derivation and also complicates the analytical approach of changing variables to move into a rotating frame of reference.



Thus motivated, we defer to Appendix~\ref{app:UpperSAF} our derivation of the   upper-projection SAF, and here we summarize our main findings.
By repeating steps similar to those presented in Secs.~\ref{sec:SAFOnGraphs} and \ref{sec:HLK SAF}, we obtain an upper-projection SAF given by
\begin{equation}
\label{eq:UpperSAFChiText}
   \frac{||L_2^{[-],\dagger} \boldsymbol{\omega}^{[+]}||^2_2}{N^{[+]}}+ \chi,
\end{equation}
where $\chi = ||\mathbf{v}||^2_2 /N^{[+]} -\left( \ \overline{L_2^{[-]\dagger} \boldsymbol{\omega}^{[+]}} \ \right)^2$ and $\mathbf{v}$ is the part of the phase-locked states $\boldsymbol{\theta}^{[+],*}$ that is in the kernel of matrix $L_2^{[-]}$.
Here, the extra term $\chi$ significantly increases the difficulty for developing computational algorithms to optimize synchronization for the upper projection.
However, in Appendix \ref{app:UpperSAF2} we provide numerical evidence that $\chi$ is usually very small and has a minimal effect on solutions to optimization problems that we will introduce in the next section. 
Thus, for the remainder of this paper, we will neglect $\chi$ and focus on  optimizing upper-projected HLK dynamics through
\begin{align}
    \label{eq:UpperSAF}
    J^{[+]}(\boldsymbol{\omega}^{[+]}, L_2^{[-]})  &= N^{[+]-1} ||L_2^{[-],\dag} 
    \boldsymbol{\omega}^{[+]}||_2^2 \nonumber\\
    &= \frac{1}{N^{[+]}} \sum_{n:\  \lambda^{[+]}_n \neq 0} \frac{(\boldsymbol{\omega}^{[+],T} {\bf v}_{[+]}^{(n)})^2}{\lambda_n^{[+]2}}
\end{align}
for eigenvector/eigenvalues pairs ${\bf v}_{[+]}^{(n)}, \lambda_n^{[+]}$ of $L_2^{[-]}$. Here, the summation is over the non-zero eigenvalues for $L_2^{[-]}=B_2^TB_2$.  It follows that Eq.~\eqref{eq:ProjectedOrderParameters} can be approximated for large $r^{[+]}$ by
\begin{equation}
    \label{eq:Up SAF order parameter}
    r^{[+]} \approx 1-J^{[+]}(\boldsymbol{\omega}^{[+]}, L_2^{[-]})/2\sigma_{[+]}^2.
\end{equation}

\section{Optimization Experiments}
\label{sec:SAFoptimization}

The two SAFs, $J^{[\pm]}$, quantitatively measure the ability for synchronization to occur in the lower- and upper-dimensional projections of the HLK dynamics. They provide objective functions for optimization, which we   highlight in this section by posing and solving
a few different optimization problems.
In Sec.~\ref{sec:Reordering}, we maximize $r^{[\pm]}$ for a fixed SC and a variable frequency vector $\boldsymbol{\omega}$ (with entries that must be optimally assigned to the 1-simplices). 
In Sec.~\ref{sec:Opt Dimensional Constraint}, we pose and solve a spectrally constrained problem, studying the effects of the balancing parameter $\delta$ and the role Hodge subspaces and homology. 
In Sec. \ref{sec:Noise}, we extend this optimization problem by adding low-level noise to disallow the use of orthogonalization for the construction of an optimal solution.

\subsection{Optimal Allocation of Oscillator Frequencies} \label{sec:Reordering}
Consider a scenario in which one has a fixed,  given SC along with a collection of oscillators, each with a particular natural frequency $\omega_i$. We would like to place these oscillators on the edges of the SC in a way that optimizes the Kuramoto order parameters $r^{[\pm]}$ for the  projected HLK model.  While it may be possible to directly simulate all possible arrangements using a brute-force method for small SCs, this approach is computationally expensive and infeasible for large SCs (i.e., since the number of combinatorial possibilities grow factorially with SC size).  

Here, we develop numerically efficient algorithms to solve this combinatorial optimization problem by combining our newly developed SAFs with a Markov chain Monte Carlo (MCMC) approach. 
(We note that a similar problem and numerical algorithms were introduced for the original graph-based SAF  \cite{skardal2014optimal}.)
First, we randomly assign frequencies $\omega_i$ to the 1-simplices in the SC. 
Next, rather than simulate the HLK dynamics and numerically observe $r^{[\pm]}$, we instead analytically calculate the associated SAFs $J^{[\pm]}$. We note that this is a computationally efficient calculation that requires 
only the vector of frequencies and spectral decomposition of the SCs' Hodge Laplacians (which can be calculated once, prior to the MCMC iterations).
At each step of the MCMC, we select two 1-simplices uniformly at random and consider swapping their associated frequencies. The new SAFs are then calculated, and the proposed swap is accepted if the new SAFs have decreased (i.e., the associated estimates for $r^{[\pm]}$ have increased).

Obviously, some frequency swaps may introduce a different affect on $J^{[-]}$ versus $J^{[+]}$ (e.g., increasing one while decreasing the other). 
Thus, we must choose between optimizing one SAF or the other, or we can try to simultaneously optimize both.
To this end, 
we aim to optimize both SAFs simultaneously by aiming to maximize the   mean of their associated order parameters
\begin{align}
    \label{eq:mean order parameter}
    \overline{r} &= \frac{r^{[-]}+r^{[+]}}{2}\\
    &\approx 1- \frac{J^{[-]}(\boldsymbol{\omega}^{[-]}, L_0)}{4\sigma_{[-]}^2} - \frac{J^{[+]}(\boldsymbol{\omega}^{[+]}, L_2^{[-]})}{4\sigma_{[+]}^2}.
    \label{eq:mean order parameter2}
\end{align}
That is, maximizing $\overline{r}$ can be achieved  by minimizing a weighted average of the SAFs with weights ${1}/{\sigma_{[\pm]} ^2}$ that are determined by their associated coupling strengths: $\sigma_{[\pm]} =\sigma(1\pm \delta) $.  

\begin{figure*}[t!]
    \includegraphics[width=\linewidth]{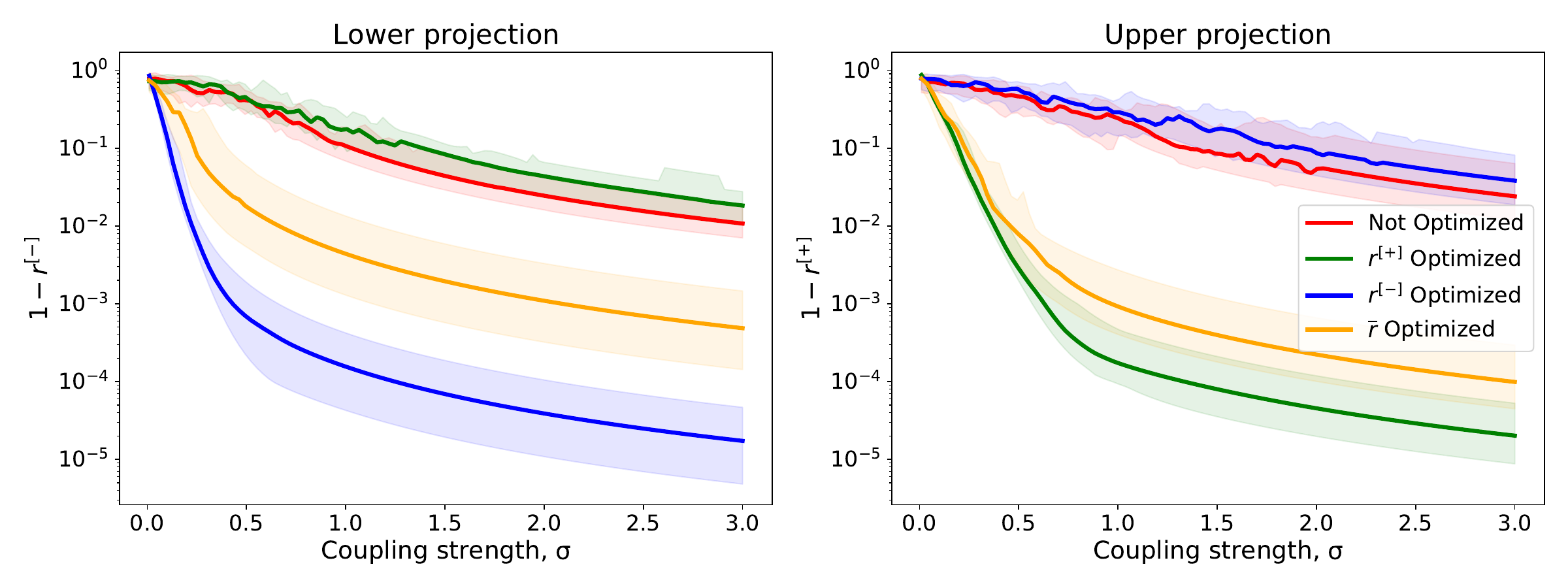}
    \label{fig:optimization simulations}
    \vspace{-.5cm}
  \caption{{\bf Optimization of lower- and upper-projected HLK model}. We plot both synchronization order parameters $r^{[-]}$ (left) and $r^{[+]}$ (right) while aiming to maximize each of them separately, or maximize them simultaneously through their mean $\overline{r}$. These $r^{[\pm]}$ values are plotted versus coupling strength $\sigma_{[\pm]}=\sigma$ (or equivalently, $\delta=0$) and compared to $r^{[\pm]}$ for random systems that have not been optimized. 
  Observe that optimizing $\overline{r}$ significantly improves synchronization for both projections.
  }
\end{figure*}

In Fig.~\ref{fig:optimization simulations}, we study the performance of our MCMC algorithm for maximizing phase synchronization for three optimization goals: maximizing $r^{[-]}$, $r^{[+]}$ or $\overline{r}$. The left and right panels depict $1-r^{[-]}$ and $1-r^{[+]}$, respectively, for optimal frequency vectors $\boldsymbol{\omega}$ that were obtained using MCMC for different coupling strengths $\sigma$. We set $\delta=0$ so that $\sigma_{[\pm]} =\sigma$.
The SC used for this experiment is discussed and visualized in Appendix~\ref{appendix:small SC}.  For each value of $\sigma$, 50 trials were simulated to identify the average order parameters.  The bounds of the shaded regions indicate the 10th and 90th percentiles across those trials. Each trial had the same frequency vector $\omega$, but different random initial phases. In each trial, 50,000 steps were used for the MCMC process to identify the optimal allocation of frequencies to the 1-simplices in the SC.

Observe in Fig.~\ref{fig:optimization simulations} that when we optimize just a single order parameter (i.e., either $r^{[-]}$ or $r^{[+]}$), that targeted order parameter can be greatly improved, while the other one does not change much (i.e., it is simular to values for the  non-optimized system). Importantly, when $\overline{r}$ is optimized, then both order parameters $r^{[\pm]}$ significantly increase (although not quite as much as for the case of optimizing each individual projection independently).

\subsection{Optimization with a Spectral Constraint}
\label{sec:Opt Dimensional Constraint}
%

For graph-coupled oscillators, the original SAF has a spectral form suggesting that synchronization can be enhanced by aligning the frequency vector $\boldsymbol{\omega}$ with eigenvectors associated with large eigenvalues of the graph Laplacian $L_0$   \cite{skardal2019synchronization}. Because the newly derived SAFs, $J^{[\pm]}$, given by Eqs.~\eqref{eq:LowerSAF} and \eqref{eq:UpperSAF}, have a similar spectral form, they allow for a similar approach, although each utilizes a projected frequency vector and a different type of Laplacian matrix. We'll now take a closer look on how the frequency vector can be constructed in a way that maximizes the mean of order parameter, $\overline{r}$, optimizing for synchronization in both projections simultaneously. We will propose an optimization problem in which $\boldsymbol{\omega}$ is spread evenly across $k$ eigenvectors of the Hodge Laplacian $L_1$.  We will then investigate which eigenvectors lead to better synchronization, and which subspace(s) of $L_1$ contain the optimal $\boldsymbol{\omega}$, as well as how that can change by altering the balancing parameter $\delta$.

Note we can aim to maximize $r^{[-]}$ by aligning the lower-projected frequency vector $\boldsymbol{\omega}^{[-]}=B_1\boldsymbol{\omega}$ with the eigenvectors ${\bf v}_{[-]}^{(i)}$ associated with large eigenvalues $\lambda_i^{[-]}$ for matrix $L_0=B_1B_1^T$. At the same time, we  can  can aim to maximize $r^{[+]}$ by aligning the upper-projected frequency vector $\boldsymbol{\omega}^{[+]}=B_2^T\boldsymbol{\omega}$ with the eigenvectors ${\bf v}_{[+]}^{(i)}$ associated with large eigenvalues $\lambda_i^{[+]}$  for matrix $L_2^{[-]}=B_2^TB_2$. 
To simultaneously optimize both projected dynamics, we instead seek to maximize $\overline{r}$.
Using these  spectral decompositions, Eq.~\eqref{eq:mean order parameter2} yields 
\begin{equation}
\label{eq:rbar}
    \overline{r} \approx 1  -
     \frac{1}{4 N^{[-]} (1-\delta)^2 \sigma^2} \sum_i \frac{(\boldsymbol{\omega}^{[-]T} {\bf v}_{[-]}^{(i)})^2}{\lambda_i^{[-]2}}- \frac{1}{ 4 N^{[+]} (1+\delta)^2 \sigma^2 } \sum_i \frac{(\boldsymbol{\omega}^{[+]T} {\bf v}_{[+]}^{(i)})^2}{\lambda_i^{[+]2}}.
\end{equation} 
%
%


Motivated by this spectral approximation, we   propose a family of spectrally constrained optimization problems in which we aim to maximize $\overline{r}$, for which we provide analytical and numerical solutions.
That is, we propose an optimization problem in which $\boldsymbol{\omega}$ is spread evenly across $k$ eigenvectors of the Hodge Laplacian $L_1$.  For a SC with $N$ edges and an integer $k \in [1, N]$,  we consider a frequency vector that must take the spectral form  
\begin{equation}
\label{eq:dimensional constriant}
    \boldsymbol{\omega} = \frac{1}{\sqrt{k}} \sum_{i=1}^k {\bf v}_i,
\end{equation}
where $\{{\bf v}_i\}$ are normalized eigenvectors of $L_1$ and the scaling
$1/ \sqrt{k}$   ensures normalization:  $||\boldsymbol{\omega}||_2=1$.  
We will analytically solve such problems using the newly derived SAFs.
We will also use  the Hodge Decomposition to predict and study the extent to which an optimal solution vector $\boldsymbol{\omega}$  is contained within, or spreads across, the gradient, curl, and harmonic subspaces associated with  $L_1$. Note that this approach subsequently reveals how SC homology has a role in shaping  higher-order networks that are optimized for synchronization under HLK oscillator dynamics.

To proceed, recall from Sec.~\ref{sec:HodgeLaplacian} that the 1-Hodge Laplacian is given by $L_1  = B_1^TB_1 + B_{2}B_{2}^T$, and we highlight that the eigenvalues $\lambda_i^{[-]}$ of matrix $L_0=B_1 B_1^T$ are identical to as those of $B_1^TB_1$, and the same is true for the eigenvalues $\lambda_i^{[+]}$ of matrices $L_2^{[-]}=B_2^T B_2$ and $B_2 B_2^T$. (It's also worth noting that these eigenvalues are also equivalent to the square of the singular values for $B_1$ and $B_2$, respectively.) 
Each nonzero eigenvalue of $L_1$ is also an eigenvector of $L_0$ or $L_2^{[-]}$, and the corresponding projection of an eigenvector of $L_1$ is also an eigenvalue of $L_0$ or $L_2^{[-]}$ (depending on where the corresponding eigenvalue came from). The lower projection of an eigenvector of $L_1$ that does not correspond to an eigenvalue of $L_0$ is ${\bf 0}$, and similarly for the upper case. We need to be careful here, as while ${\bf v}_i$ is normalized and its lower (or upper) projection is an eigenvector of $L_0$ (or $L_2^{[-]}$), this projection $B_1 {\bf v}_i$ (or $B_2^T {\bf v}_i$) is not normalized, and so does not replace ${\bf v}_{[-]}^{(i)}$ (or ${\bf v}_{[+]}^{(i)}$) in Eq.~\ref{eq:rbar}.  Instead, if $({\bf v}_i, \lambda_i)$ is an eigenvector/eigenvalue pair of $L_1$ associated with the eigenvector ${\bf v}_{[-]}^{(j)}$ of $L_0$, then when we project downward, $B_1 {\bf v}_i = \sqrt{\lambda_i}{\bf v}_{[-]}^{(j)}$, where $\sqrt{\lambda_i}$ is the corresponding singular value of $B_1$, and similarly for the upper projection.  

With these insights, it follows that  the projected frequency vectors take the form 
\begin{align}\label{eq:ProjectedFrequencyDown}
    \boldsymbol{\omega}^{[\pm]} = \frac{1}{\sqrt{k}} \sum_i^{k^{[\pm]}} \sqrt{\lambda^{[\pm]}_i }{\bf v}_{[\pm]}^{(i)},
\end{align} 
where 
$k^{[-]}$ is defined as the number of eigenvectors of $L_1$ that are included in $\boldsymbol{\omega}$ and which correspond  to an eigenvector of $L_0$ (and similarly, $k^{[+]}$ is the number of eigenvectors associated with $L_2^{[-]}$).
%
It follows that 
$(\boldsymbol{\omega}^{[\pm]T} {\bf v}_{[\pm]}^{(i)})^2 = \lambda_i/k$ if ${\bf v}_i$ is an eigenvector of $L_1$ that is used to construct $\boldsymbol{\omega}$, and $(\boldsymbol{\omega}^{[\pm]T} {\bf v}_{[\pm]}^{(i)})^2 =0$ otherwise. Furthermore, Eq.~\eqref{eq:rbar} takes the form
\begin{equation}
\label{eq:asdasd}
    \overline{r} \approx 1- \frac{1}{4 \sigma^2} \left ( \sum_{\mu_i^- \in \mathcal{S}^-} \frac{(\boldsymbol{\omega}^{T} {\bf v}_{i})^2}{\mu_i^-} + \sum_{\mu_i^+ \in \mathcal{S}^+} \frac{(\boldsymbol{\omega}^{T} {\bf v}_{i})^2}{\mu_i^+} \right ),
\end{equation}
where we have introduced new parameters
\begin{equation}
    \mu_i^{\pm}= N^{[\pm]} (1\pm\delta)^2 \lambda_i^{[\pm]}
\end{equation} 
as well as the multisets $\mathcal{S}^- = \{\mu_i^-\}$ and $\mathcal{S}^+ = \{\mu_i^+\}$. (Recall that a multiset is a set that possibly allows for repeated entries.)


%
This spectrally constrained optimization problem requires one to solve the combinatorial problem of choosing which eigenvectors of $L_1$ to use to construct  $\boldsymbol{\omega}$. 
Because the numerators in Eq.~\eqref{eq:asdasd} will all be similar regardless of which eigenvectors are chosen (i.e., either $0$ or $1/k$), we seek to maximize $\overline{r}$ by choosing the $k$ eigenvectors of $L_1$ associated with the largest denominators; that is, the terms associated with the largest $\mu_i^{[\pm]}$ values.
%
We assume that both sets of eigenvalues are indexed in descending order so the $\mu^{\pm}_i$ increase with their index.

To formalize our selection process, we
%
consider the gradient, curl, and harmonic subspaces of $L_1 $  from Eq.~\eqref{eq:HodgeDecomp}, which has the positive eigenvalues
$\{\lambda_n^{[\pm]}\}$ as well as zero eigenvalues, the number of which is equal to the dimension of the harmonic subspace. Recall from Sec.~\ref{sec:HodgeLaplacian} that the dimension of the harmonic subspace is equal to the first Betti number, $\beta_1$, and is determined by the SC's  homology. We define $\mathcal{H}$ to be the set  of eigenvectors spanning the harmonic subspace, which is of size   $\beta_1$.  
As discussed in Sec.~\ref{sec:phase locking}, the harmonic projection of $\boldsymbol{\omega}$ has no effect on   synchronization for the projected dynamics, and therefore it does not contribute to either SAF, $J^{[\pm]}$. For example, if $\boldsymbol{\omega}$ lies entirely within the harmonic space,  $\text{span}(\mathcal{H})$, then it is in the kernel of both $B_1$ and $B^T_{2}$, implying $\boldsymbol{\omega}^{[\pm]}   = \mathbf{0}$ and
${r}^{[\pm]} = \overline{r}=1$. That is,
perfect synchronization occurs for both the lower- and upper-projected HLK dynamics, and we can therefore interpret the harmonic subspace as a ``safe haven'' to assign frequency heterogeneity without diminishing synchronization under HLK dynamics. 
Thus, for any $k\leq \beta_1$, we can construct an optimal frequency vector $\boldsymbol{\omega}^*$ by choosing any $k$ eigenvectors from  the set $\mathcal{H}$. (The solution is not unique unless $k=\beta_1$.)

If we were to consider constructing a simplicial complex that allows for the strongest synchronization, it is tempting to maximize the size of the harmonic subspace.  This allows for more ``room'' to ``hide'' the frequency vector in components that do not affect the synchronization dynamics.  However, as the dimension of the harmonic subspace is the same as the first Betti number, increasing this dimension without increasing the number of oscillators (edges) is done by removing 2-simplices (triangles).  Maximizing the dimension of this subspace would also remove all triangles from the simplicial complex, eliminating all upper-dimensional interactions entirely. 

To construct an optimal solution for larger values of $k$, we first select the  $\beta_1$ eigenvectors from  $\mathcal{H}$. Next, we must choose $k-\beta_1$ additional nonzero eigenvectors that are associated with $L_0$ and/or $L_2^{[-]}$.  To do so, recall our definitions of  nonzero$\mathcal{S}^-$ and $\mathcal{S}^+$ for Eq.~\eqref{eq:asdasd}. We can think of their elements $\mu_i^{\pm}$ as measures for how easily synchronization can occur if $\boldsymbol{\omega} $ aligns with an eigenvector ${\bf v}_i$.  Let $\mathcal{S}_n$ be the multiset  of the $n$ largest elements of $\mathcal{S}^- \cup \mathcal{S}^+$, and let $\mathcal{V}_n$ be their associated eigenvectors of $L_1$---that is, for each 
$\mu_i^{\pm}= N^{[\pm]} (1\pm\delta)^2 \lambda_i^{[\pm]}$, 
it will follow that $\lambda^{[\pm]}_i$ is an eigenvalue of $L_1$, and   $\mathcal{V}_n$ consists of their  $n$  associated eigenvectors. 
%
Letting $n=k-\beta_1$ so that $\mathcal{S}_{n}=\mathcal{S}_{k-\beta_1}$ and $\mathcal{V}_{n}=\mathcal{V}_{k-\beta_1}$, we now  present a general SAF-based approximate solution to this spectrally constrained optimization problem: 
\begin{equation}
    \label{eq:optimal frequency}
    \boldsymbol{\omega}^* = \begin{cases}
        \frac{1}{\sqrt{k}} \displaystyle\sum_{\substack{i=1 \\ {\bf v}_h^{(i)} \in \mathcal{H}} }^k {\bf v}_h^{(i)}, \ \ k \leq \beta_1 \\ \\
        \frac{1}{\sqrt{k}} \left (\displaystyle\sum_{{\bf v}_h \in \mathcal{H}} {\bf v}_h + \sum_{{\bf v}_i \in \mathcal{V}_{k-\beta_1}} {\bf v}_i \right ), \ \  k > \beta_1.
    \end{cases}
\end{equation}
Here, we have used a subscript `h' to allude to the harmonic subspace, $\text{span}(\mathcal{H})$.
%
It is important to note here that $\mathcal{S}_{k-\beta_1}$ will always be unique, but the vector elements of $\mathcal{V}_{k-\beta_1}$ may not be unique, e.g., if there are repeated eigenvalues (or if $\mu_i^{+} = \mu_j^{-}$  for some $i$ and $j$, which can  occur since they both vary with $\delta$). In the case of repeated $\mu_i^{\pm}$ values, it is therefore possible to have a non-unique solution to the optimal frequency vector $\boldsymbol{\omega}^*$; however, all such solutions will 
%
yield the same  SAF-approximated value for $\overline{r}$.

\begin{figure*}[b!]
    \centering
    \includegraphics[width=.85\linewidth]{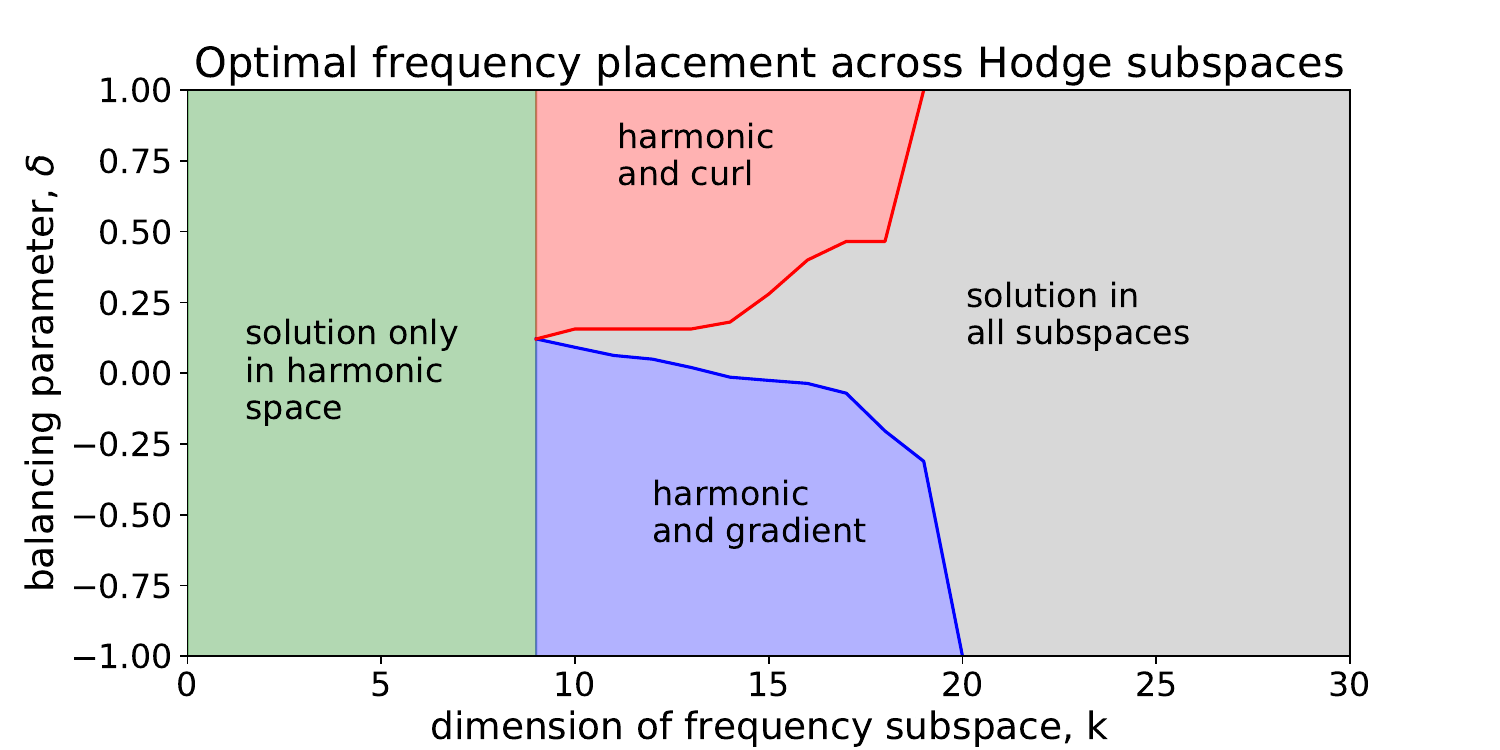}    
  \caption{{\bf Bifurcations for the alignment of  $\boldsymbol{\omega}^*$  with the Hodge subspaces}. Optimal frequency vector $\boldsymbol{\omega}^*$ is constructed using $k$ eigenvectors ${\bf v}_i$ that lie with in the gradient, curl and harmonic subspaces for $L_1$. Depending on $k$ and $\delta$ (which varies the relative onset of synchronization for   lower- and upper-projections and  tunes the $\mu_i^{\pm}$ values), $\boldsymbol{\omega}^*$ can be contained in one, two, or all three of these Hodge subspaces. The SC is discussed in Appendix~\ref{appendix:small SC}, and observe   that $\beta_1=9$. 
    %
  }
  \label{fig:Bifurcation}
\end{figure*}

In Fig.~\ref{fig:Bifurcation}, we study bifurcations
for how
$\boldsymbol{\omega}^*$ spreads across the three   Hodge subspaces, 
depending on $k$ and $\delta$ (which tunes the $\{\mu_j^\pm\}$ values), 
We have already discussed why $\boldsymbol{\omega}^*\in\text{span}(\mathcal{H})$ for any $k\le \beta_1$. To understand why $\boldsymbol{\omega}^*$ might be orthogonal to either the curl or gradient subspace for $k>\beta_1$, we consider the case when $\delta\to1$, which implies that 
%
$\sigma^{[-]}\to0$ and $\mu_i^{[-]}\to0$. In this limit,  the optimal solution vector would select eigenvectors associated with the larger $\mu_i^{[+]}$ values,  whose eigenvectors lie within the curl subspace. A similar argument can be made for the gradient subspace when  $\delta\to-1$.

Perhaps counterintuitively, this means that the solution $\boldsymbol{\omega}^*$ given by Eq.~\eqref{eq:optimal frequency} first focuses on enhancing synchronization for the projection that is easier to synchronize. This occurs  because if $\boldsymbol{\omega}^*$ is orthogonal to a Hodge subspace, then its associated projected dynamics can achieve perfect synchronization. That is, $r^{[-]}=1$ if $\boldsymbol{\omega}^*$ is orthogonal to the gradient subspace, since orthogonality causes the projected frequencies to become identical, $\boldsymbol{\omega}^{[-]}=0$. Similarly, $r^{[+]}=1$ if $\boldsymbol{\omega}^*$ is orthogonal to the curl subspace. (It was these properties that led to our identification of the harmonic subspace as a `safe haven.')
If $k>\beta_1$, but $k$ is not too large, then the optimal frequency vector can take advantage of this situation,  and $\boldsymbol{\omega}^*$ will be orthogonal to one Hodge subspace, if possible. (This may not be possible for  intermediate values of $\delta$, since the $\mu_i^{[-]}$ and $\mu_i^{[+]}$ values will be interleaved.) Also, if $k$ is too large, then $\boldsymbol{\omega}^*$ must spread across all three Hodge subspaces because its dimension $k$ makes the solution simply too big to fit within one of these subspaces. (For example, $\boldsymbol{\omega}^*$ cannot be contained within the harmonic and curl subspace if $k$ is larger than the dimension of these combined subspaces.)
%
%
%

In Fig.~\ref{fig:Bifurcation}, red and blue curves identify 
critical values of $\delta$ where $\boldsymbol{\omega}^*$ undergoes these
bifurcations. That is,  $\boldsymbol{\omega}^*$ is orthogonal to the gradient subspace whenever $\delta>\delta^{+}(k)$ (red curve), and
it is orthogonal to the curl subspace whenever $\delta<\delta^{-}(k)$ (blue curve), where
\begin{equation}\label{eq:crit_d}
    \delta^\pm(k) = \left({\sqrt{\gamma_{k-\beta_1}^{[\pm]}}- 1}\right) ~/~\left({\sqrt{\gamma_{k-\beta_1}^{[\pm]}}+ 1}\right)
\end{equation}
and $\gamma_{i}^{[-]} = \frac{N^{[-]} \lambda^{[-]}_i}{N^{[+]} \lambda^{[+]}_1}$ and $\gamma_{i}^{[+]} = \frac{N^{[-]} \lambda^{[-]}_1}{N^{[+]} \lambda^{[+]}_i}$.
%
%
This derivation follows from considering that the set $\mathcal{V}_n$ will select   eigenvectors associated with the largest $\mu_i^{[\pm]}$ values, and that the $\mu_i^{[+]}$ values will increase  as $\delta$ increases (while the $\mu_i^{[-]}$ values  decrease). Thus, for any $k>\beta_1$ we can ask, how large $\delta$ needs to be to guarantee that at least $k$  $\mu_i^{[+]}$ values   are strictly larger than all of the $\mu_i^{[-]}$ values.
We denote this value $\delta^+(k)$, and we  obtained Eq.~\eqref{eq:crit_d} by setting $\mu^{[+]}_{k-\beta_1} = \mu^{[-]}_1$ 
and  solving for $\delta$. We obtained $\delta^-(k)$ using a similar technique.


Next, we discuss the optimal parameter $\overline{r}^*$ that corresponds to the optimal frequency vector $\boldsymbol{\omega}^*$.  Note that all of the numerators  in Eq.~\eqref{eq:asdasd} are  either $\frac{1}{k}$ or $0$, depending on whether or not the associated eigenvector is used to construct $\boldsymbol{\omega}^*$. We therefore substitute this solution form to obtain a SAF-approximated maximum order parameter 
%
\begin{equation}\label{eq:r2}
    \overline{r}^*  \approx 1 - \frac{1}{4k \sigma^2}\sum_{\mu_i^{\pm} \in \mathcal{S}_{k-\beta_1}}\frac{1}{\mu_i^{\pm}}.
\end{equation}

\begin{figure*}[t!]
    \centering
    \includegraphics[width=.9\linewidth]{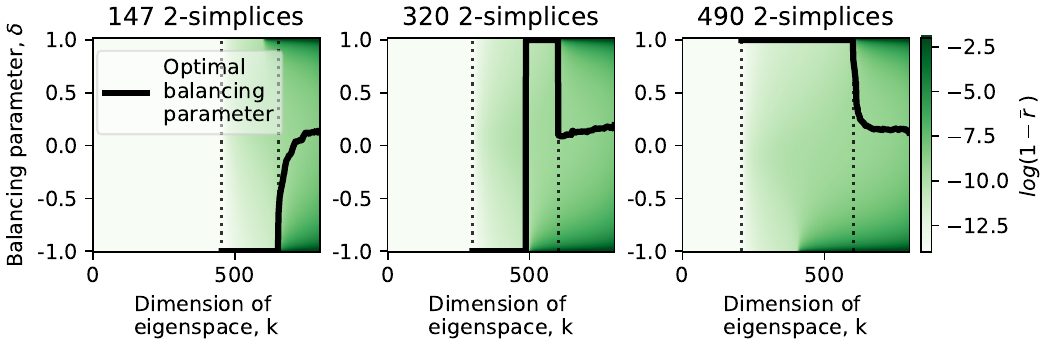}
    \vspace{-.2cm}
    \caption{{\bf Optimal order parameter $\overline{r}^*$ with optimal balancing $\hat{\delta}$}. Heat maps illustrate $\overline{r}^*$ across the $(k,\delta)$ parameter space for three SCs. Each SC has the same 1-skeleton (i.e., the same 0- and 1-simplices) but a varying number of 2-simplices. The black curve identifies the optimal balancing parameter $\delta$ versus $k$. Vertical dotted lines predict where the solution frequency vector $\boldsymbol{\omega}$ is expected to undergo bifurcations w.r.t. which Hodge subspaces it lies within.
    %
}\label{fig:Noiseless examples}
\end{figure*}

In Fig.~\ref{fig:Noiseless examples}, we depict $\overline{r}$ given by Eq.~\eqref{eq:r2} as a function of $k$ and $\delta$ using heat maps. The three panels depict results for three different SCs, which all have the same 0- and 1-simplices but different numbers of 2-simplices. See Appendix~\ref{appendix:small SC} for  discussion. In each panel, a black curve identifies the balancing parameter $\hat{\delta} \equiv \text{argmax}_\delta \overline{r}^* $ that maximizes   $\overline{r}$ for each $k$. 
Note that $\hat{\delta}$ is not plotted only for $k\le\beta_1$ (the left-most, vertical dotted line in each panel),   where $\overline{r}=1$ and $\delta$ has no effect on $\boldsymbol{\omega}^*$. The right-most, vertical dotted line in each panel indicates either (left panel) the sum of $\beta_1$ and the gradient subspace's dimension, or (middle and right panels) the sum of $\beta_1$ and the curl subspace's dimension.

We conclude this section by making a few insights from  Fig.~\ref{fig:Noiseless examples}. Focusing on the left panel, observe for $k>\beta_1$ but $k \approx \beta_1$ that $\hat{\delta}$ is exactly $\delta=-1$, implying that $\overline{r}$ is maximized by  allocating the entire available coupling  (i.e., $2\sigma = \sigma_{[+]} + \sigma_{[-]} $) to $\sigma_{[-]}$ so that the 1-simplices are only coupled through lower-adjacent interactions. In this case,   we seek to maximize $r^{[-]}$, and   $r^{[+]}=1$ occurs because $\boldsymbol{\omega}^*$ is orthogonal to the curl subspace. Then around $k\approx 750$, $\hat{\delta}$ diverges from $\delta=-1$ (which occurs when $k$ is too large and $\boldsymbol{\omega}^*$ is no longer orthogonal to the curl subspace).

Turning our attention to the center panel, observe that as the number of 2-simplices in the SC increases, $\beta_1$ becomes smaller (i.e., the left-most, vertical dotted line shifts leftward). Also, the black curve (i.e., $\hat{\delta}$) now exhibits a discontinuous jump from $\delta=-1$ to $\delta=1$ near $k\approx 500$. That is, as  $k$ surpasses $\delta\approx 500$,   $\overline{r}$ is now maximized by making $\boldsymbol{\omega}^*$ orthogonal to the gradient subspace, and by allocating the entire available coupling  to $\sigma_{[+]}$ so that the 1-simplices are only coupled through upper-adjacent interactions.  The discrete jump from $\delta=-1$  to $\delta=1$ highlights that the system can still design $\boldsymbol{\omega}^*$ to be orthogonal to one of the Hodge subspaces, but it can no longer do this for the stronger-synchronizing projection (i.e., the lower projection). Then, as $k$ surpasses approximately $600$, a second bifurcation occurs, since $k$ is so large that $\boldsymbol{\omega}^*$ can no longer be contained within just 2 of the Hodge subspaces. 

Finally, the right panel in Fig.~\ref{fig:Noiseless examples} shows a situation where $\beta_1$ continues to become smaller. Also, it is now the case the $\delta=-1$ is never optimal.
This change for $\hat{\delta}$ should be expected, since the addition of 2-simplices into the SC shrinks the dimensionality of the harmonic subspace and grows the curl subspace. The growing curl subspace involves the matrix $B_2$ becoming larger and larger, thereby introducing more  eigenvalues  $\lambda_i^{[+]}$ for $L_2^{[-]}$ (and in particular, larger-valued ones). As we mentioned earlier, the nature of our solution $\boldsymbol{\omega}^*$ is that as $k$ surpasses $\beta_1$, the coupling resources are allocated to maximizing synchronization for the projection that is easier to synchronize (while  orthogonality is used to achieve perfect synchronization in the projection that is harder to synchronize). As the number of 2-simplices in the SC increases, these roles switch for the two projected HLK systems.


\subsection{Optimization with Orthogonality Disallowed}\label{sec:Noise}
%
The analytical solution $\boldsymbol{\omega}^*$ given by Eq.~\eqref{eq:optimal frequency} is constructed by selecting   eigenvectors   associated with large denominator terms, $\mu_i^{\pm}$.
From another perspective, the solution is built by make $\boldsymbol{\omega}^*$ to be orthogonal to the eigenvectors   associated with small denominators. (The name synchrony `alignment' function emphasizes the alignment with certain eigenvectors, but for some cases avoidance from others can be equally, if not more, important.) Also, perfect orthogonality may be infeasible for some scenarios.
 
In this section, we extend the optimization problem from Sec.~\ref{sec:Opt Dimensional Constraint} with low-level noise to disallow perfect orthogonalization when constructing a solution $\boldsymbol{\omega}^*$. 
That is, we seek to maximize $\overline{r}$ for a family of frequency vector that must take the form 
\begin{equation}
    \label{eq:noisy dimensional constraint}
    \boldsymbol{\omega} = \sqrt{\frac{1-\epsilon}{k}}\sum_{i=1}^k {\bf v}_i + \sqrt{\frac{\epsilon}{N-k}} \sum_{i = k+1}^N {\bf v}_i,
\end{equation}
where $\mathcal{V}=\{ {\bf v}_i\}$ is an eigenbasis for $\mathbb{R}^N$ using eigenvectors for $L_1$.
The coefficients are defined to ensure that $|| \boldsymbol{\omega}||_2 = 1$ for any values of $k > 0$ and $\epsilon \in [0, 1]$. 
This is a more realistic problem, as any amount of uncertainty in the frequency vector could result in an unexpected small contribution from any eigenvector (including those associated with small $\mu_i^{[\pm]}$, which might significantly inhibit synchronization).  
Also, note in this limiting cases with $\delta = \pm 1$, synchronization would be impossible for one of the two projections.


\begin{figure*}[b!]
    \includegraphics[width=\linewidth]{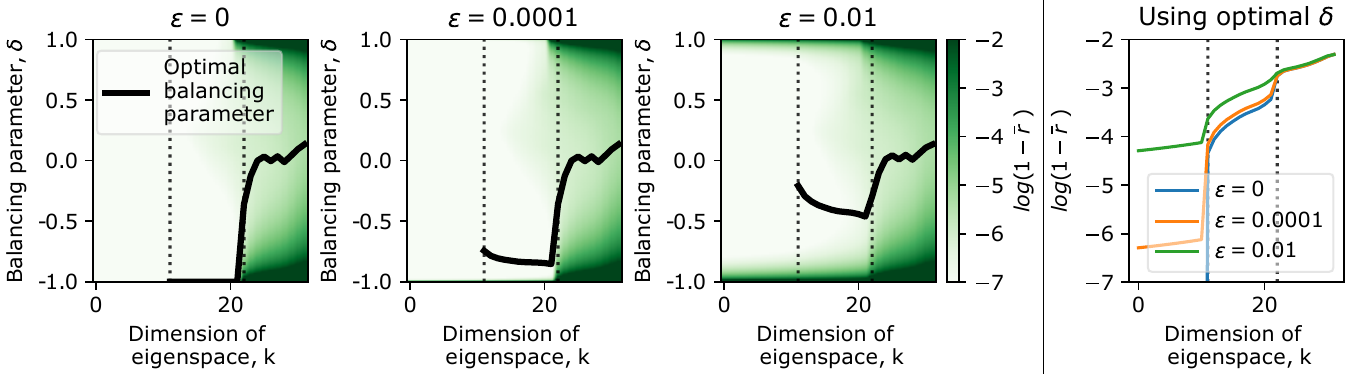}
    \vspace{-.4cm}
    \label{fig:Noise Plot}
  \caption{{\bf Optimization of $\overline{r}$ when orthogonalization is disallowed.} The three left-most panels are constructed similarly to those in Fig \ref{fig:Noiseless examples}, although  we focus on the SC with the fewest 2-simplices. We plot $\overline{r}$ for an optimal solution frequency vector constrained to the form given by Eq.~\eqref{eq:noisy dimensional constraint} with three choices for $\epsilon$.  Black curves depict the optimal balancing parameter $\hat{\delta} \equiv \text{argmax}_\delta \overline{r} $ for each $k$.
  The right-most panel show the $\overline{r}$ value associated with $\hat{\delta}$ versus $k$.
}
\end{figure*}

To solve this problem, we must still decide which $k$ vectors ${\bf v}_i$ should receive the larger coefficients, and it turns out that these can be selected identically as before. That is, we first select  up to $\beta_1$ eigenvectors associated with the harmonic subspace. Then if $k>\beta_1$, we we select the eigenvectors associated with  the largest $\mu_i^{\pm}$ values (noting that these will differ as $\delta$ is varied).
%
%

In left   panels of Fig \ref{fig:Noise Plot}, we use heat maps to visualize $\overline{r}^*$ using the same SC that was studied in the first panel Fig.~\ref{fig:Noiseless examples}. Panels 1-3 were reflect   three different choices for $\epsilon$. Note that the overall behavior of $\overline{r}^*$ does not change much; however,  the optimal balancing parameter $\hat{\delta}$ versus $k$ significantly  changes (black curves).
These results are somewhat expected, since subspace orthogonality has been strategically disallowed for this experiment. For example, it is never possible for the optimal solution $\boldsymbol{\omega}^*$ to have $\delta=\pm 1$---the choice would make synchronization impossible for one of the two projections. Note that even when  $\epsilon>0$ is very small, maximizing $\overline{r}$ involves using a more balanced choice for $\delta$. 
Thus, when orthogonality is not allowed, effort is required to simultaneously synchronize both projections (i.e., both $r^{[-]}$ and $r^{[+]}$), and orthogonalizing $\boldsymbol{\omega}^*$ away from either the curl or gradient subspace is not available as a design mechanism.

In the right-most panel of Fig \ref{fig:Noise Plot}, we depict the $\overline{r}$ values associated with  with each $\hat{\delta}$ curve, as a function of $k$. Three curves indicate the three choices of $\epsilon$, and in all panels, vertical dotted lines indicate $\beta_1$ (left) and the sum of $\beta_1$ and the gradient subspace's dimension (right).
Observe that increasing $\epsilon$ decreases synchronization, especially when $k < \beta_1$. However, once $k$ is large enough so that $ \boldsymbol{\omega}$ has large coefficients across  all three subspaces, $\epsilon$'s affect on $\overline{r}$ essentially vanishes.
Therefore, even though this optimization problem intentionally prevents perfect orthogonalization,  it is still very insightful to understand how the optimal frequency vector $\boldsymbol{\omega}^*$ aligns (or misaligns) with the different Hodge subspaces.


\section{Conclusion}\label{sec:conclusion}

Recently, there has been growing interest to generalize  dynamics to the setting of higher-order networks using the mathematical models such as simplicial complexes \cite{bick2023higher}. This provides an opportunity  for algebraic topology to extend into new application areas.  It has been used, for example, to study  dynamical systems including consensus   \cite{ziegler2022balanced}, synchronization \cite{nurisso2024unified,millanexplosive} and random walks \cite{schaub2020random} as well as develop algorithms for signal processing \cite{barbarossa2020topological, calmon2023dirac} and machine learning \cite{roddenberry2019hodgenet}.
Such work often utilizes the study of Hodge Laplacians and their associated theory (e.g., Hodge decompositions). Nevertheless, there remains limited optimization theory and tools for this  emerging research field.
%
%
Thus motivated, here we have extended the Synchrony Alignment Function (SAF) \cite{skardal2014optimal,skardal2016optimal,taylor2016synchronization} for optimizing synchronization for graph-coupled heterogeneous oscillators to the setting of simplicial complexes. By extending the SAF framework to the Hodge Laplacian Kuramoto (HLK) model \cite{millanexplosive,nurisso2024unified}, our work complements other SAF extensions to simplicial complexes \cite{skardal2021higher} that formulate oscillator dynamics in a different way that does not directly involve the Hodge Laplacian or algebraic topology.

Here, we have optimized the HLK model in which oscillators are assigned to 1-simplices (i.e., edges) and  synchronization occurs for the two associated  systems in which the HLK dynamics is projected onto the 0-simplices (i.e., nodes) and 2-simplices (i.e., filled-in triangles). We have also introduced a small extension to the HLK model by introducing a balancing parameter $\delta$ as a way to tune the relative onset of synchronization for these two  projections (which aligns this work with our prior introduction of the balanced Hodge Laplacian  \cite{ziegler2022balanced}). 
%
We derived two SAF functions, $J^{[\pm]}$, that can be readily used to optimize synchronization (i.e., maximize Kuramoto order parameters $r^{[\pm]}$) for the lower- and upper-projections of the HLK dynamics. Interestingly, these HLK SAFs were derived to have a similar spectral-theory form allowing a system's synchronizability to be interpreted using matrix spectral theory. (We note that $J^{[+]}$ was found to require an additional approximation due to added complications that arise when considering SC's 2-dimensional boundary matrices and homology.)
That is, the newly derived SAFs allow us to quantify and optimize extent of synchronization based solely on a simplicial complex's structure (as encoded in its balanced Hodge Laplacians) and the oscillators' natural frequencies. 
Similar to the original SAF, our approach utilizes linear approximation theory for the strong-synchronization regime, although it's worth highlighting that the optimization framework's applicability extends well beyond this regime.

We highlighted the broad utility of the HLK SAF framework though several constrained optimization problems.  In one, we considered the optimal placement of natural frequencies on a given simplicial complex. We showed that  synchronization for both the lower- and upper-dimensional can be simultaneously enhanced by optimizing their two SAFs simultaneously.  
A second family of optimization problems was also studied in which the frequency vector is ``spectrally'' constrained to a linear combination of $k$ eigenvectors for the Hodge Laplacian $L_1$.  We studied how the synchrony-optimizing eigenvectors were related to the three Hodge subspaces, and constructed bifurcation diagrams to characterize when the optimal solution lies in one, two, or all three of these subspaces.
These findings integrate ideas from combinatorial optimization theory, algebraic topology, and dynamical systems, thereby pointing to an exciting new emerging field for interdisciplinary applied math.

\appendix
\section{Simplicial Complexes used for Demonstration and Experiment}\label{appendix:small SC}

\begin{figure}[h!]
    \label{fig:exampleSCs}
    \includegraphics[width=\linewidth]{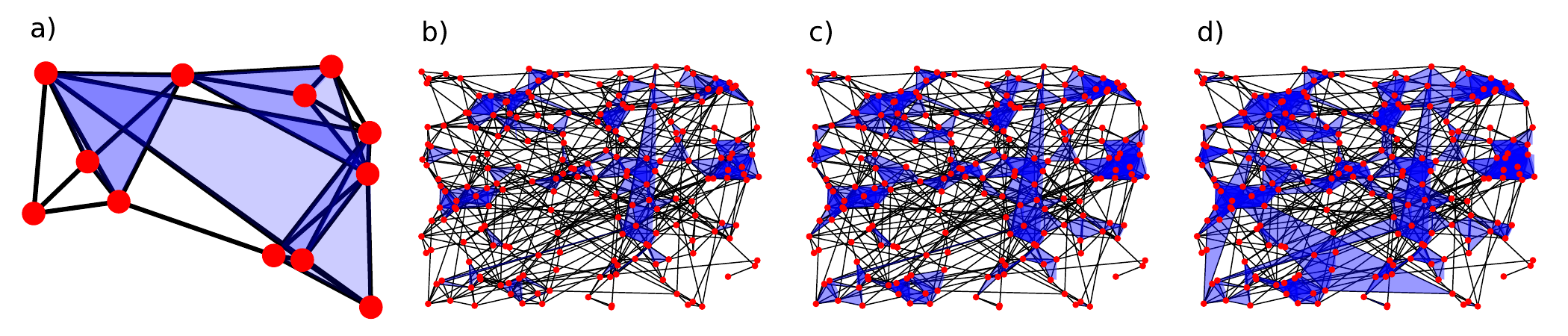}
    \vspace{-.5cm}
    \caption{{\bf Simplicial Complexes studied throughout our paper.} (a) SC 1 was used for Figures \ref{fig:SynchSimulations}-\ref{fig:Bifurcation} and \ref{fig:Noise Plot}  (b)--(d) The remaining SCs were  used for Fig.~\ref{fig:Noiseless examples}.}
\end{figure}

In this paper, we make frequent use of the following simplicial complexes.
The SC shown in Fig.~\ref{fig:exampleSCs}(a) was used for Figs.~\ref{fig:SynchSimulations}--\ref{fig:Bifurcation} and \ref{fig:Noise Plot}. It has twelve 0-simplices, thirty 1-simplices, and ten 2-simplices.  It was created by first assigning 12 0-simplices  random $(x, y)$ coordinates.  Then, 1-simplices were randomly created one at a time by either adding them between the pair of 0-simplices that are closest together (which occurred with probability 0.8), or otherwise creating them between a pair of unconnected 0-simplices selected uniformly at random. Each unfilled 3-cycle was then chosen to be filled with a 2-simplex with a probability of 0.5.

The three SCs shown in Figs.~\ref{fig:exampleSCs}(b)--(d) were used for our numerical experiments that were described in
Fig.~\ref{fig:Noiseless examples}.
Each of these has an identical set of two-hundred 0-simplices and eight-hundred  1-simplices, but they have different numbers of 2-simplices.  They were created using the same method as SC 1.  To create 2-simplices, we considered all the unfilled 3-cycles and filled them in with a 2-simplex with one of three probabilities: $0.2, 0.4,$ and $0.6$. 
This was done in a way such that all of the 2-simplices in an SC associated with a smaller probability also exist in the SC(s) associated with larger probabilities.

\section{Derivation of SAF for Upper-Projected HLK Model}
\label{app:UpperSAF}
Here, we derive  the upper-projected HLK SAF using steps that are very similar to those presented  in Sec.~\ref{sec:HLK SAF} for the lower-projected HLK SAF. That is, we seek to maximize the order parameter $r^{[+]}$ given by Eq.~\eqref{eq:ProjectedOrderParameters} for dynamics given by Eq.\eqref{eq:KuramotoDecoupledUp}.

When the system exhibits a phase-locked state, it follows that
$\dot{\boldsymbol{\theta}}^{[+]} = \mathbf{0}$, and the projected oscillators reach a steady phase $\boldsymbol{\theta}^{[+],*} \approx \mathbf{0}$. This state is guaranteed to exist  for large enough coupling strength $\sigma_{[+]}$ \cite{nurisso2024unified}, and it follows that Eq.~\eqref{eq:KuramotoDecoupledUp} reduces to 
\begin{equation}
    \sigma_{[+]} L_2^{[-]} \sin{(\boldsymbol{\theta}^{[+],*})}= \boldsymbol{\omega}^{[+]}.
\end{equation}
Because $\boldsymbol{\omega}^{[+]}=B_2^T \boldsymbol{\omega} \in \text{Im}(B_2^T) = \text{Im}(L_2^{[-]})$, we continue by multiplying both sides by the Moore-Penrose pseudoinverse $\sigma_{[+]}^{-1}L_2^{[-]\dag}$
and simplify to obtain
\begin{equation}
    \label{eq:UpSAF1}
    \sin(\boldsymbol{\theta}^{[+],*}) = \sigma_{[+]}^{-1}L_2^{[-]\dag} \boldsymbol{\omega}^{[+]} + \mathbf{v},
\end{equation}
where $\mathbf{v} \in \text{ker}(L_2^{[-],\dag}) = \text{ker}(L_2^{[-]}) = \ker(B_2)$ is the part of $\sin(\boldsymbol{\theta}^{[+],*})$ in this kernel.  Unlike for the derivation of the lower-dimensional SAF (which relies on the kernel  $\text{ker}(L_0)   = \ker(B_0^T)=\text{span}({\bf 1})$, that is, assuming that the SC is connected), this higher-dimensional kernel is more complicated and can be empty or have large dimension.

Again, it is not necessarily true that $\boldsymbol{\theta}^{[+],*} \approx \mathbf{0}$, but rather, 
$\boldsymbol{\theta}^{[+],*}  \approx \mathbf{0}\ (\text{mod} \ 2\pi)$.  To remedy this, we consider the vector 
$\boldsymbol{\theta}^{[+],*}-2\pi\mathbf{z} \approx \mathbf{0}$ for some  $\mathbf{z} \in \mathbb{Z}^{N^{[+]}}$, which is defined as the vector of integers such that  $\boldsymbol{\theta}^{[+],*}-2\pi\mathbf{z} \in [-\pi, \pi)$.  
After this substitution,  Eq.~\eqref{eq:UpSAF1} can be linearized to obtain 
\begin{equation}
    \label{eq:UpSAF2}
    \boldsymbol{\theta}^{[+],*}-2\pi \mathbf{z} = \sigma_{[+]}^{-1}L_2^{[-]\dagger}\boldsymbol{\omega}^{[+]} + \mathbf{v}.
\end{equation}
As long as there are no 3-simplices, this kernel is the harmonic space for $L_2$, and so this kernel represents the SC's 2-dimensional holes  (i.e., cavities). A basis for this kernel consists of a vector for each cavity. Each contains zeros except for the entries associated with the 2-simplices that define the cavity's boundary (which are assigned $\pm 1$  values depending on their orientations).

To proceed, we take the average of Eq.~\eqref{eq:UpSAF2}; that is, we multiply both sides of by $\mathbf{1}/N_{[+]}$ to obtain 
\begin{equation}
    \label{eq:UpSAF3} 
    \overline{\boldsymbol{\theta}^{[+],*}-2\pi \mathbf{z}} = \overline{\sigma_{[+]}^{-1}L_2^{[-]\dagger} \boldsymbol{\omega}^{[+]}} + 0.
\end{equation}
The term $\overline{\mathbf{v}}$ becomes $0$, because $\mathbf{1} \in \text{Im}(B_2^T)$. However, we cannot cancel out the other term on the right-hand side, since $\mathbf{1} \not \in \ker(B_2)$.  
%
Next, we combine Eqs.~\eqref{eq:UpSAF2} and \eqref{eq:UpSAF3} to obtain 
\begin{equation}
    \boldsymbol{\theta}^{[+],*}-2\pi \mathbf{z}  -\overline{\boldsymbol{\theta}^{[+],*}-2\pi \mathbf{z}} = \sigma_{[+]}^{-1}L_2^{[-]\dagger}\boldsymbol{\omega}^{[+]} +\mathbf{v}-\overline{\sigma_{[+]}^{-1}L_2^{[-]\dagger}\boldsymbol{\omega}^{[+]}} .
\end{equation}
Without changing the order parameter, $r^{[+]}$, 
next we change the reference frame $\boldsymbol{\theta}^{[+],*} \rightarrow \boldsymbol{\theta}^{[+],*}-2\pi \mathbf{z}$ to yield 
\begin{equation}
    \boldsymbol{\theta}^{[+],*} -\overline{\boldsymbol{\theta}^{[+],*}} \mathbf{1}= \sigma_{[+]}^{-1}L_2^{[-]\dagger}\boldsymbol{\omega}^{[+]} + \mathbf{v} -\overline{\sigma_{[+]}^{-1}L_2^{[-]\dagger}\boldsymbol{\omega}^{[+]}} \mathbf{1}.
\end{equation}
Finally, we can define and solve for an associated variance order parameter (recall Sec.~\ref{sec:SAFOnGraphs}) given by
\begin{equation}
\begin{aligned}
    R^{[+]} &= 1-\sigma_{\boldsymbol{\theta}^{[+],*}}^2/2 \\
    &= 1-\frac{1}{2N^{[+]}}||\mathbf{\boldsymbol{\theta}^{[+],*}} - \overline{\boldsymbol{\theta}^{[+],*}}\mathbf{1}||^2 \\
    &\approx 1-\frac{1}{2N^{[+]}}||\sigma_{[+]}^{-1}L_2^{[-]\dagger}\boldsymbol{\omega}^{[+]} + \mathbf{v} -\overline{\sigma_{[+]}^{-1}L_2^{[-]\dagger}\boldsymbol{\omega}^{[+]} } \mathbf{1}||^2.
\end{aligned}
\end{equation}

It follows for $r^{[+]} \approx 1$ that we can approximate
\begin{equation}
    \label{eq:Up SAF order parameter Appendix}
    r^{[+]} \approx 1-J^{[+]}(\boldsymbol{\omega}^{[+]}, L_2^{[-]})/2\sigma_{[+]}^2,
\end{equation} 
where we have defined a SAF for the upper-projected HLK model:
\begin{align}
    J^{[+]}(\boldsymbol{\omega}^{[+]}, L_2^{[-]}) &= \frac{||L_2^{[-]\dagger}\boldsymbol{\omega}^{[+]} +\mathbf{v}-\overline{L_2^{[-]\dagger}\boldsymbol{\omega}^{[+]}} \mathbf{1}||^2_2}{N^{[+]}} \label{eq:FullUpperSAF}\\
     &= \frac{||L_2^{[-]\dagger}\boldsymbol{\omega}^{[+]}||^2_2}{N^{[+]}}+ \chi .
     \label{eq:FullUpperSAFChi}
\end{align}
This simplification is obtained by expanding the norm and using the facts that $\mathbf{v} \in \text{ker}(L_2^{[-]\dag})$ and $\overline{\mathbf{v}} = 0$.

Here, we have introduced a correction term
\begin{equation} \label{eq:Chi}
    \chi =  ||\mathbf{v}||^2_2 /N^{[+]} - \left( \ \overline{L_2^{[-]\dagger}\boldsymbol{\omega}^{[+]} } \ \right)^2 ,
\end{equation}
%
%
which did not arise for the original SAF, $J$, nor the lower-projection SAF,  $J^{[-]}$. We also note  for SCs that contain no 3-simplices and have empty 2-dimensional homology, the first term in $\chi$ is zero because ${\bf v}$ is in the kernel of $L_2^{\text{down}}$, which is zero. 
%
Importantly, the existence of $\chi$ can make the optimization of $J^{[+]}$ and $r^{[+]}$ significantly much more difficult.
In the next section, we show that
$\chi $ can be considered to be a small correction term  with little impact on $J^{[+]}$ (at least this was observed for  the numerical experiments discussed herein). Thus motivated, we have neglected the $\chi$ term when developing numerical and theoretical solutions to the optimization problems posed in Sec.~\ref{sec:SAFoptimization}.
%

\section{Numerical Evidence Showing that  $\chi$ has a Small Effect}
\label{app:UpperSAF2}
Here, we provide numerical evidence  to show that in general, $\chi$ is relatively small compared to $J^{[+]}$ and can be reasonably neglected. %
One benefit is that if one ignores $\chi$, then 
Eq.~\eqref{eq:FullUpperSAFChi} 
allows for an interpretable  understanding of $J^{[+]}$ through eigenvalues and eigenvectors of $L_2^{[-]}$.  That is, a frequency vector that aligns strongly with the dominant eigenvectors of $L_2^{[-]}$ yields a small SAF and strong synchronization.  
It follows that the numerical optimization of $J^{[+]}$ becomes much easier.

\begin{figure*}[h]
    \centering
    \includegraphics[width=.6\linewidth]{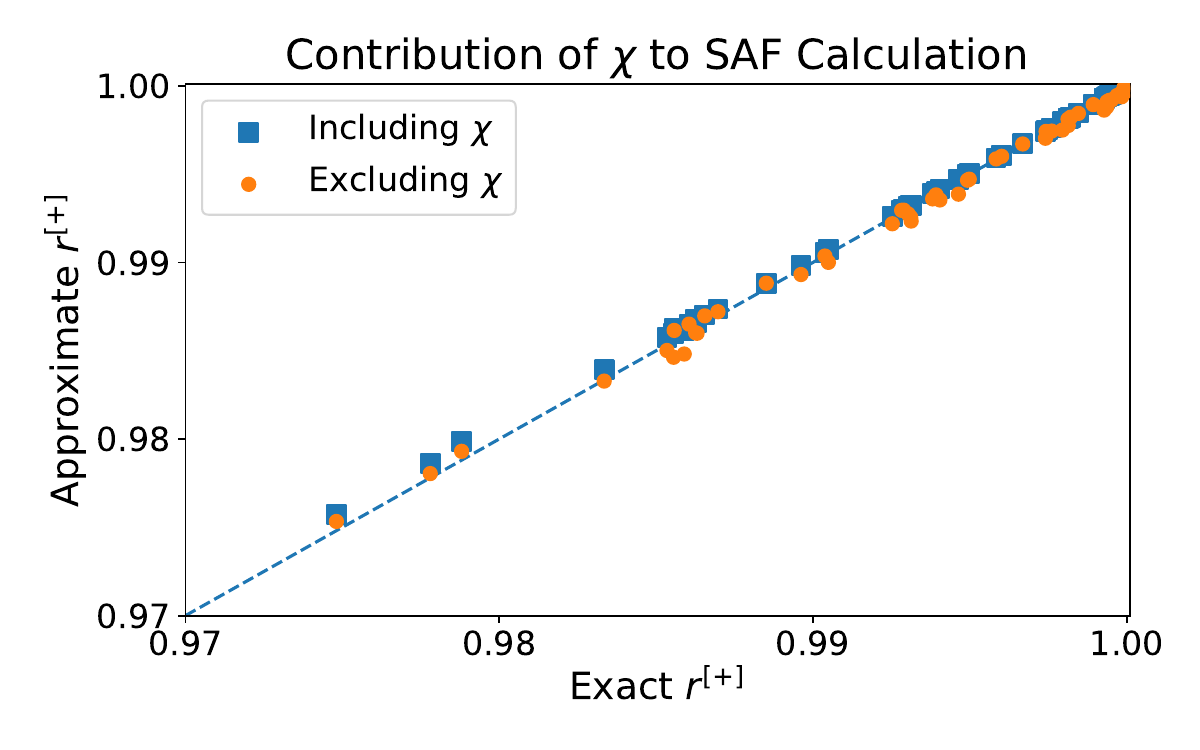}
    \vspace{-.5cm}
    \caption{{\bf Evidence showing that $\chi$ has small impact.} We compare measure values of the order parameter $r^{[+]}$ given by Eq.~\eqref{eq:ProjectedOrderParameters} versus theoretical predictions obtained using the upper-dimensional SAF $J^{[+]}$ while either including or ignoring the correction term $\chi$.    
    While this term is not wholly negligible, it does not provide a large contribution to $J^{[+]}$, and we find that it can be ignored while having minimal impact. } 
  \label{fig:Chi Plot}
\end{figure*}

To provide numerical support for neglecting  $\chi$, in Fig.~\ref{fig:Chi Plot} we compare direct measurements for $r^{[+]}$ to estimated values using Eq.~\eqref{eq:Up SAF order parameter Appendix} with either $\chi$ included or omitted. The data points represent  independent experiments for 100 different SCs, which we constructed using the generation method discussed in Appendix \ref{appendix:small SC} 
Each SC was created using a random choice of parameters, and we focused our attention to SC having at least  fifty 0-simplices and three-hundred 1-simplices.  

Observe in Fig.~\ref{fig:Chi Plot}  that while it is not quite negligible in general, $\chi$ is small enough to ignore for applications that do not require great precision.  The optimization problems that we consider in this paper are not sensitive enough for $\chi$ to significantly change the resulting optimized networks.  Thus, we choose to neglect this term in our calculations in favor of greatly simplifying how we theoretically and numerically solve herein.

\section*{Acknowledgments}
This work is supported by the U.S. National Science Foundation under Grant No. DMS-2401276. Any opinions, findings, and conclusions or recommendations expressed in this material are those of the authors and do not necessarily reflect the views of the National Science Foundation.

\bibliographystyle{siamplain}
\bibliography{references}

\end{document}